# The extraordinary story of Sinann — the inspirational figure who gave her name to Ireland longest river — and how she arose Ireland's resilient female icons

Ralph Kenna[1,2], Chris Thompson[3], Isolde Ó Brolcháin Carmody[3], Benjamin Dwyer[4], Daniel Curley[5], Mike McCarthy[5], Nicola Bowes[6], Pádraig MacCarron[2,7], Thierry Platini[1,2] and Joseph Yose[1,2]

1. Centre for Fluid and Complex Systems, Coventry University, Coventry CV1 5FB, United Kingdom
2. L4 Collaboration & Doctoral College for the Statistical Physics of Complex Systems, Leipzig-Lorraine-Lviv-Coventry, Europe
3. Story Archaeology, Carrick on Shannon, Leitrim, N41D982, Ireland
4. Middlesex University, Faculty of Arts & Creative Industries, The Burroughs, London NW4 4BT
5. Rathcroghan Visitor Centre, Tulsk, Roscommon, Ireland
6. Celtic Eye, Eastwell, Cappataggle, Ballinasloe, Co Galway H53NV09, Ireland
7. MACSI, Department of Mathematics and Statistics, University of Limerick, Limerick V94 T9PX, Ireland

* Corresponding author: Ralph Kenna, kenna.ralph@gmail.com

**Abstract:** When local authorities recently selected a neoclassical male "river god head" of colonial origin to represent Ireland's longest river, it was welcomed as "harking back to Irish mythology". The council, local historians and townsfolk were unaware that in Irish mythology the figure associated with the river is a woman — not a man. Her name is Sinann, and she had been written out of Ireland's national iconography after centuries of colonial destruction of Gaelic heritage. When mathematical investigation into Irish mythology brought Sinann's story to the people via local media, reaction was immediate. Street performances in support of Sinann were backed by letters in newspapers and a petition signed by hundreds of people demanding education about their heritage - and respect for women. How did this happen and what is the story behind Sinann, her diminution, and her rising?

Here we recount how modern-day mathematics exposed the colonial imposter as stemming from the Ossianic controversy 250 years ago. We also discuss how a Victorian-era translation of her story from the original Irish turned what was an inspirational creation myth into a tale of a disobedient girl seeking knowledge to which she had no entitlement. This flawed narrative entered the public domain — in encyclopaedias, on websites, and even in academic literature. Here we cast off colonial baggage to present a new, more accurate translation of Sinann's story alongside a new interpretation — more amenable to a receptive public seeking enlightenment. Inspired by Sinann's rising, we also use artistic measures to invoke other Irish resilient female icons to rise up and take their rightful place in Irish iconography.

# 0. Introduction

When the *Westmeath Independent* newspaper announced the selection of a neoclassical male "river god head" to represent the town of Athlone and its relation to Ireland's longest river, the Shannon, one of the authors of this paper was aghast. Because of his work on Irish mythology Ralph Kenna was well aware the river was female in indigenous culture (see Appendix A). Kenna saw a double insult: misappropriation of Irish mythology for colonial aims and misogynistic overwriting of a woman's place as a powerful Irish icon. A letter to the local newspaper came fast and the ensuing protest was fierce; although uninformed as to the impending statue's iconographic meaning, the letter resonated with public instinct, and they too were soon aghast. And from that sprung the reawakening of one of Ireland's greatest mythological icons; Sinann (or Sionann, pronounced "Shun-awn" or "Shin-un") arose to claim her place as the only figure associated with the River Shannon. Newspaper letters, radio and TV interviews, street actions, an open letter and petition all spread awareness of Sinann as the legitimate figurehead of the river. An arts project engaged Ireland's local and vast diasporic communities and Sinann irrevocably staked her claim as sole authority of the river.

So how did Westmeath County Council make such a blunder in selecting a statue that the public would quickly come to despise? What was the colonial mindset behind their decision? And who is Sinann, what is her story and why is she so relevant today? That is the subject of this paper. Its authors, a selection of physicists, mathematicians, mythologists, musicians, historians, archaeologists, and artists echo the universality of support for Sinann's re-coming at a time of global crisis.

In the next section, we recount the tale of public opposition to the neo-colonial statue and outpouring of support for Sinann. In Section II we document a funded arts project that enabled the public to engage deeply with her as a significant figure in Irish mythology. In Section III we provide contextualisation for Eugene O'Curry's Victorian-era translation of Sinann's story, trimmed down versions of which became widespread. We also give a new and very different translation. Free from colonial and Victorian baggage, we hope the new translation will enter the public domain following this paper. In Section IV we explain the deep roots of a colonial mindset that still persists in Ireland today and that ultimately lies behind the council's artistic selection. In Section V, a participatory research project is used to counter this mindset by evoking Ireland's resilient female icons in the form a high-profile arts project. We summarise this long paper in Section VI and give our hopes and dreams for the future. These are that this paper goes some way to arousing the goddesses of Ireland to ensure they will never again be forgotten or overwritten by colonial, neoclassical or misogyny that, unlike Sinann, are long past their time.

# I. The Rise of Sinann

The story of the rise of Sinann is indeed an extraordinary one. Here we tell of how she rose from obscurity to belovedness. And this came about because of the unusual combination of mathematics and mythology. This came together in 2010 under a *Leverhulme Trust* funded project called Maths Meets Myths (MMM). MMM used network science to study the configurations of characters in various texts and focus from the outset was on Irish mythology. The many epic narratives studied under MMM included the Mythological and Finn Cycles in which Sinann makes at least a cameo appearance. Another study was of the epic poems of Ossian [Yose et al., 2016]. The upshot of that study was that, as is widely accepted in many quarters, Ossian misappropriated much of its material from the Finn Cycle when it was written down in the 18$^{th}$ century. Ossian caused "one of the most famous literary controversies of all time" as Ireland's scholars rallied to defend the country's heritage, under severe threat as it was in colonial times. It is extraordinary, therefore, how that altercation was echoed when in 2019, as an Ossianic style statue started to make its way to Athlone — a town in central Ireland that sits on the River Shannon. In this section we recount this tale as it unfolded; we document how MMM re-ignited love for Sinann and crushed the Ossianic statue in the minds of the townsfolk and Ireland's vast diaspora.

**I.A. The River God of the Shannon?**

In 2018, Westmeath County Council set aside €60,000 for a new sculpture to "reflect the enhancement" of the town of Athlone. Proposals for the sculpture were to address[1]:

- The arts in Athlone town's urban context
- The location of the artwork in relation to its environs, in particular the River Shannon and the cycleway
- The heritage, memory and environment of Athlone.

The deadline for tender proposals was August 9 of that year. Eight submissions were duly received, and three pieces were shortlisted by a "select group" comprising (a) an artist representative, (b) a retired librarian, (c) a member of the executive, (d) the Westmeath Arts Officer and (e) a member of Athlone Municipal District.[2]

On 17 January 2019, the *Westmeath Independent* reported that scale models of the three pieces (Figure 2.1) would be on display in Athlone Civic Centre until 28 January. Comments from the public were to be fed back to the selection panel who would "make the final decision shortly afterwards". However, these public observations "would not be counted."

---

[1] "Sculpture to reflect town centre facelift", *Westmeath Independent*, 26 July 2018; https://www.westmeathindependent.ie/2018/07/26/sculpture-to-reflect-town-centre-facelift/

[2] Deirdre Verney, "Views sought on eye-catching Church Street sculpture ideas", *Westmeath Independent*, 17 January 2019; https://www.westmeathindependent.ie/2019/01/17/views-sought-on-eye-catching-church-street-sculpture-ideas

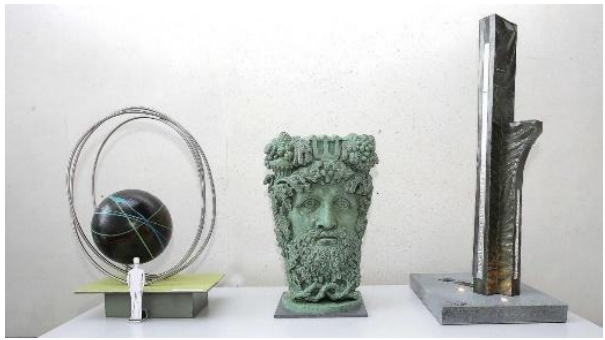

**Figure 2.1:** Three options for the sculpture as displayed in Athlone Civic Centre in 2019. The central piece, called Mask of the Shannon, was the winner. Credit: *Westmeath Independent*, 23 January 2019, https://www.westmeathindependent.ie/news/roundup/articles/2019/01/23/4168133-three-options-for-athlone-town-centre-sculpture-on-display/

The following week, the *Westmeath Independent* gave more information on the three pieces. The description of the eventual winner (the middle of the three in Figure 2.1) reads[3]:

> The Mask of the Shannon is a 180 degree bronze mask, depicting an adaptation of Edward Smyth's Shannon keystone on the Custom House. Fierce and proud, this mask-icon of the River Shannon is laden with fruits of the river's basin. The back of the sculpture presents a modern circuitry effect, indicative of Athlone's key location, point of crossing and centrality.

The Custom House referred to here is a neoclassical 18th century building in Dublin, whose original purpose was to collect duties on exports to the British Empire.

On 4 May 2019 the outcome of the winning sculpture was announced by Gearoid O'Brien, a highly regarded local historian and columnist for the *Westmeath Independent*. O'Brien wrote[4]:

> I was delighted to see the piece of sculpture chosen…When the three short-listed pieces were exhibited for public viewing in Athlone Civic Centre, I have to admit that this piece, featuring the head of the River God of the Shannon, was my personal favourite.

O'Brien went on to describe how the same iconography featured on old Irish £50 notes and how "River God" heads were first devised by a sculptor called Edward Smyth who was born in Ireland in 1749. He also described how they were "commissioned from Smyth by James Gandon as ornamental features for Dublin's Custom House." Following a brief discussion of Smyth's and Gandon's lives, O'Brien continued:

---

[3] "Three options for Athlone town centre sculpture on display", *Westmeath Independent*, 23 January 2019; https://www.westmeathindependent.ie/news/roundup/articles/2019/01/23/4168133-three-options-for-athlone-town-centre-sculpture-on-display/

[4] Gearoid O'Brien, "Welcome news on Church Street sculpture", *Westmeath Independent*, 4 May 2019.

> One of the most distinctive features of its ornamentation is the series of fourteen key-stones designed to represent thirteen of Ireland's best-known commercial rivers. It was believed that the river gods influenced the trade and commerce of the local fisheries and so they were seen as a very fitting motif for the Custom House. Each river is depicted by a unique head of a river god, and a further key-stone represents the Atlantic Ocean.

The remainder of O'Brien's article concerns the sculptor of the new piece:

> When asked about his adaptations of Smyth's heads of the river gods in a newspaper interview Rory Breslin replied: "I was passing the Custom House a few years ago and thought it was disappointing that the view of the river heads was so restricted…so I decided that I would do a new variation of them, a bit like one would have a new interpretation of a classical music."

O'Brien also commented how "the front view of the sculpture harks back to Irish mythology."

Thus, all appeared well in Athlone; the new statue was the "personal favourite" of a respected local historian and was a new interpretation of a classical theme that "harked back to Irish mythology". Moreover, it was an Irishman who devised the original for the Custom House in Dublin which thirteen others adorn. And to really be sure all is OK, they featured at national level on old Irish currency. There was no protest by any of the public, reassured as they were with the information to hand. What could go wrong?

**I.B. Sinann Awakes**

Kenna was aghast at seeing that this iconography was making its way to his hometown. Although resident in England (which is why he missed the two-week window for the local display of Figure 2.1), it was in Athlone that he was born and bred. He immediately penned an admonitory letter to the *Westmeath Independent* (25 May 2019). That letter is published in full in Appendix B1 and states (see Figure A3):

> The "river god of the Shannon" as presented in this sculpture does not in any way "hark back to Irish mythology" or, indeed, Athlone heritage. Our river was named after Sionann — the granddaughter of Lir (of the "Children of Lir" fame). This unfortunate sculpture therefore represents misappropriation of gender and aggrieves women worldwide whose names (Shannon) derive from our river deity. But the effrontery (to my mind at least) is even more than that. As Gearoid's item so ably demonstrates, the notion of a (male) "river god" of the Shannon was entirely concocted in the 18th century for the Custom House in Dublin. The river-effigies on the Custom House represent commerce and profit — hence their appearance on our national currency — and have nothing to do with our mythology. The prime position of the custom-House ornamentation is held by the crown of England, perched superior to the harp of Ireland which is surrounded by the lion and unicorn – symbols of governance in a time of subjugation. None of the symbolism, or any of its creators, has anything specific to do with Athlone.

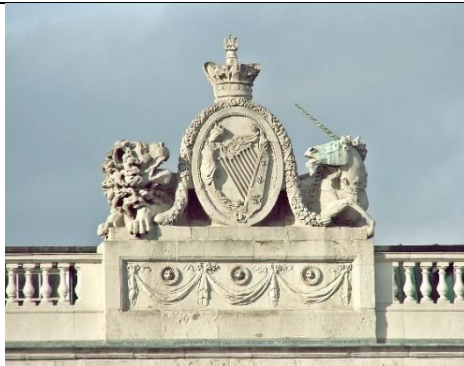

**Figure 2.2:** The crown of England is perched superior to the harp of Ireland which is surrounded by the lion and unicorn as part of the British Coat of Arms. Kenna objected to these as "symbols of governance in a time of subjugation."
Reproduced from
https://commons.wikimedia.org/wiki/File:The_Custom_House_Dublin_Crest_Irish_Harp.jpg

Kenna's letter went on to complain that while the three shortlisted pieces were exhibited for public viewing, the iconographic and factual information he gave in his letter were either "not to hand or not adequately promulgated to enable Athlonians to deliver fully informed opinions in the narrow timeframe available." Kenna called for an urgent rethink.

On 6 June, Kenna messaged Westmeath Arts Services asking about a role the council had in making the decisions, expertise available and how the selected piece matches the call. The council responded assuring the process "was conducted to a standard that adheres to the policies and procedures of Athlone Municipal District, Westmeath County Council and National Standards in the commissioning of art works. This was to be the Council's stance throughout the following story ("process is process").

On 12 July, at the invitation of local film and theatre critic Eunan Keys, Kenna was interviewed on *Athlone Community Radio* and immediately afterwards he met with the District Manager, who was a member of the selection panel. She admitted she had no expertise on iconography and was rather there to assess health and safety issues. She revealed they had hundreds of comments about the three pieces and a small majority had favoured the Mask of the Shannon (although that wasn't the basis for its selection).

**I.C. The People Rally to Sinann**

In under a week, inspired by Kenna's radio interview, members of the public Orla Donnelly and Fiona Lynam start a Facebook Public Group titled "Reclaiming Sionann for Athlone". This rapidly garnered 240 members[5] and its purpose was to start a petition in favour of Sinann.[6] The following day, Lynam was herself interviewed on *Athlone Community Radio*[7] as well as on *Midlands 103*.[8] The latter radio station had a weekly reach of 120,000 listeners and had nearly 48,000 Facebook followers. A delightful street protest was rapidly organised close to where the statue was expected to be erected (Figure 2.3). Twenty people took part, symbolically holding books on mathematics and Irish mythology. This was to acknowledge that MMM, as mentioned in Kenna's radio interview, kicked off public awareness of Sinann and of the inappropriateness of the incoming statue.

---

[5] "Reclaiming Sionann for Athlone", *Facebook*
https://www.facebook.com/groups/2604824649529222/
[6] "No to Athlone Statue," *Irish Post*, 27 July 2019.

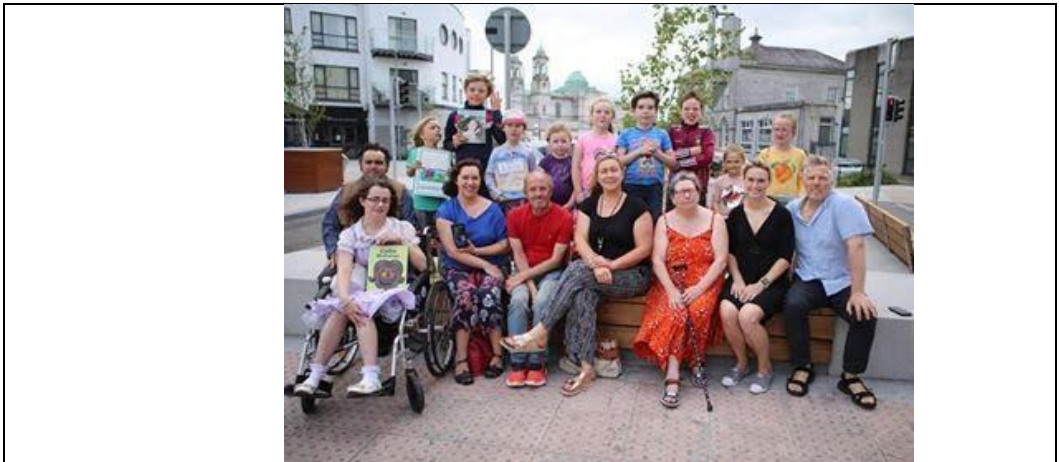

**Figure 2.3:** People at a street protest holding maths and myths books to demonstrate how MMM ignited awareness of Sinann (22 July 2019). On the 27th of July, the image was published in the *Westmeath Independent* for an item titled "Planned Athlone's artwork's 'a misappropriation of our heritage.'"

But even more powerfully, on 22 July, Lynam was interviewed on *Shannonside Radio* which had 62,000 listeners.[9] Lynam was convinced that Kenna's interview had "made the case that the sculpture is not suitable at all and bears no part in local history or mythology".

On 27 July, Kenna, Donnelly, Lynam and over 100 people of diverse backgrounds published an open letter in the *Westmeath Independent*, using the photo from Figure 2.3. The letter directly addressed the new sculpture and its adaption from the Custom House in Dublin "where it is claimed to represent a Shannon River deity":

> On behalf of hundreds of people from Athlone, people who have settled in Athlone, and Athlone's national and international diaspora, and on behalf of concerned people nationally and the world over, we object to the intended statue. It represents misappropriation of Athlone's and Ireland's heritage, an affront to Mná na hÉireann and to people who value parity.

---

[7] Athlone Community Radio, "Fiona Lynam talks about the petition to have the River God sculpture reversed" 18 July 2019;
https://www.mixcloud.com/AthloneCommunityRadio/athlone-today-fiona-lynam-talks-about-the-petition-to-have-the-river-god-sculpture-reversed/

[8] Midlands 103 "Petition to install statue of Irish goddess in Athlone [19 July 2019];
http://www.midlands103.com/news-centre/petition-to-install-statue-of-irish-goddess-in-athlone

[9] Shannonside, "Online-petition-started-installation-street-sculpture-athlone" 22 July 2019;
https://www.shannonside.ie/news/local/west-meath/online-petition-started-installation-street-sculpture-athlone

> The Custom House iconography is neo-classical and is not at all connected with Irish heritage. It claims a vacuous male god to represent our river while in Irish mythology the river is female. The name Shannon stems from Sínann and may derive from "Old Honoured One".[10] It is personified as a goddess — it is female, not male, and millennia older than the Custom House. The Metrical Dindshenchas (our native recount of the origins of placenames and traditions, committed to writing over a thousand years ago) relates how Sionann drowned at Connla's well in her pursuit of wisdom. "Rivers in Ireland…were envisaged as divine figures to whom were attributed the gift of poetical inspiration, mystical wisdom and all-encompassing knowledge." Ours is a rich and ancient mythology, one we should be proud of and take inspiration from.
>
> In contrast, the Custom House in Dublin is devoid of cultural meaning. It was designed in the 18th century by James Gandon whose world was one of privileged nobility. Beneath that world lay "the impoverished underclass…still struggling for their basic civil rights." Elitists like John Beresford, the first commissioner of revenue in Ireland who conceived the construct, insisted that they "should be kept down by a policy of unyielding repression." When Napper Tandy (co-founder of the United Irish movement of Catholic, Protestant and Dissenter) et al protested, Beresford instructed Gandon to carry on and to "laugh at the folly of the people."
>
> Thus, and as pointed out previously in public, the use of Custom House motifs for Athlone is a discordant move to say the least. The Custom house ornamentation positions Shannon mythology away from Irish sources and closer to the classics. This echoes the Ossian debacle 250 odd years ago — an attempt to supplant Ireland's mythological heroes to position a concocted narrative away from Irish mythological sources and close to the classics. It represents arrogant and brutal misappropriation of Irish mythology.
>
> The Custom House theme also misappropriates gender; although our town has no female iconography, you, our Council, deem it appropriate to commission a concocted male figure instead of the authentic female one which is steeped in our heritage and story.

The open letter is re-published fully in Appendix B1. It goes on to address the council's "process is process" defence:

> The local authority's response, claiming to have been "fully cognizant", that the decision was "a fully informed one", and that "the process is underway and cannot be reversed", more resembles Beresford's instructions to "laugh at the folly of the people" than it does Sionann's pursuit of wisdom. In her article 'The Fate of Sinann', Kiltoom-born Celtic scholar Maud Joynt (1868-1940) discusses how the "the original legend perhaps foreshadowed the dangers" of insufficient expertise. As elucidated by Professor of Irish Studies Noémie Beck, "knowledge was believed to be perilous when not handled correctly". Thus, in Sionann's story it appears that our ancestors warned of the danger of knowledge without context – the very pit that the council appears to have led us into.

The letter ends by calling "for these words to be heeded and this folly to be halted".

The letter contains extensive citations to academic references. It stands in marked contrast to the dearth of information supplied to the "hundreds" of people who engaged in the public consultation process the previous January. The newspaper simultaneously published an interview with the Council, but they only repeated their "process is process" refrain.[11] It also reported that the petition had now garnered 300 signatures.

---

[10] See however Appendix C1 where an alternative meaning is put forward.

In late July 2019, Caroline Coyle and colleagues started a second *Facebook* group called "Linking Hands Across the Bridge for Goddess Sionann." Its purpose was to organise a street event on 5 August (Figure 2.4). The event was covered in the national newspaper journal.ie where it received 29,000 views.[12] The media take up accelerated and later in August Kenna gave an 11-minute interview on *Newstalk radio*. *Newstalk* is the second most listened to station in Ireland and the Moncrieff show has 97,000 listeners.

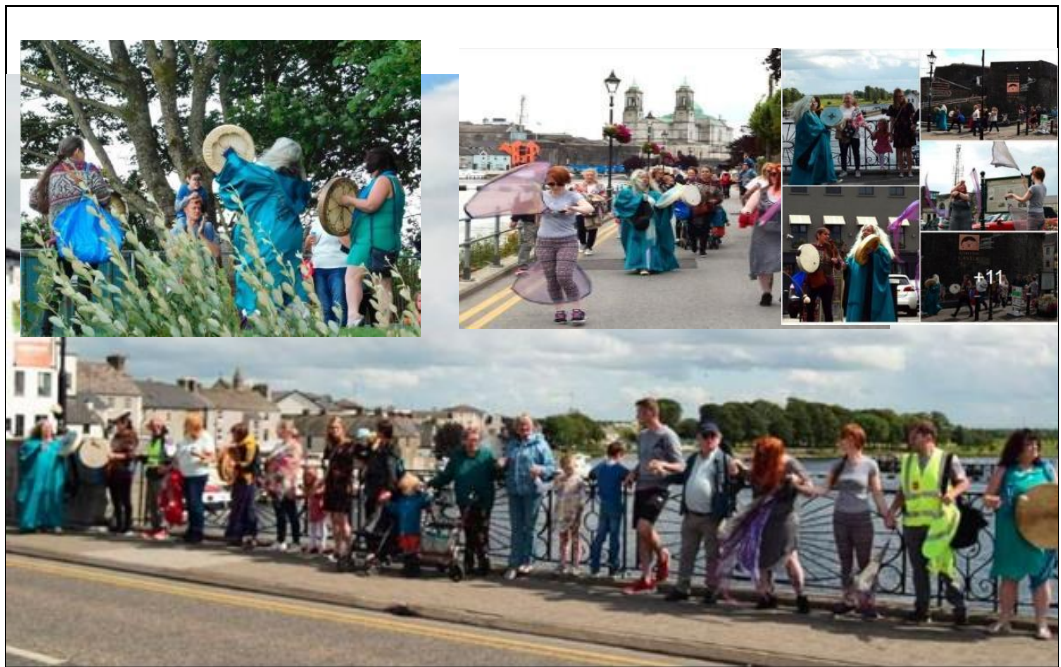

**Figure 2.4:** On 05 August 2019 the "Linking Hands Across the Bridge for Goddess Sionann" event occurs in Athlone with 50 participants and hundreds of passers-by reacting in support. Photos courtesy of John Madden and colleagues (2019).

The campaign next reached ministerial level when on 14 August, Donnelly met Minister of State Kevin Moran to discuss "possible solutions, including the idea that the council might cancel the piece" Ten days later the *Westmeath Independent* reported on national coverage and inquiries were made by the newspaper to the culture minister. The campaign impacted council meetings as well. As reported in the *Westmeath Independent*:[13] "The campaign is continuing, however, and this weekend a "Shout out for Síonnan" event is planned on a boat under Athlone's town bridge". The "collaborative community event" was described as inviting people to "embrace and celebrate Sionnan as Goddess of the Shannon through a music, river theatre and art performance." This third event took place on 1 September.

---

[11] Adrian Cusack, "Controversial artwork 'has already been commissioned' says council", *Westmeath Independent*, 27 July 2019.

[12] The Journal, "How a Shannon-side sculpture sparked a culture war in Athlone" 18 August 2019; https://www.thejournal.ie/shannon-athlone-sculpture-westmeath-river-4768807-Aug2019/

[13] Adrian Cusack, "Athlone sculpture controversy to be raised at council meeting", *Westmeath Independent*, 28 August 2019.

On 29 August 2019 the petition, by now signed by 586 people, was submitted by Lynam (the number of signatories went up to 700 shortly after submission). The petition reads as follows (it is published in full in Appendix B2):

> The "river god" statue chosen for Athlone misrepresents our native heritage and our rich culture. It would be an affront to our heritage and people to use colonial male object from Dublin to represent the Shannon and Athlone. The mythological Goddess Sionann, granddaughter of Lir, is our mythological river deity — not a concocted neo-classical god. Misappropriation of mythology and gender in a time of national subjugation is not acceptable as a modern representation of our town, nor is representation of the town on the backside of a statue. The concept of celebrating our river and our heritage is a welcome one and we call on the Council to do so by recognising our heritage – not replacing it. We call on Westmeath County Council to revoke its uninformed selection.

**I.D. The Council's Defence: Process is Process**

In September, in-council discussions and open radio interviews continued apace. In addition to this the *Athlone Advertiser* and *Athlone Topic* newspapers carried front-page items[14] and full-page reports.[15] On 7 September, the *Westmeath Independent* reported the council again as saying, "The commissioning process for the public art piece was completed in line with Westmeath County Council's Public Art Policy and included a public consultation element which gave people the opportunity to give their views on the shortlisted pieces." Thus the "process is process" mantra was maintained. There was no let-up on media coverage, however, and on 13 September Donnelly escalated the pressure through a 3-minute interview in the national TV station TG4.

Finally, on 2 October, Donnelly and Kenna had a 1.5-hour showdown in the form of a meeting with five of Athlone's six councillors, including the town mayor. The discussions were fraught with Deputy Mayor Cllr. O'Rourke especially adamant that the statue would be erected and there was to be "no turning back." At least three invitations to debate on radio were declined by the council with the mayor suggesting that the campaigners read a statement from the council instead. That statement is contained in Appendix B3. O'Rourke explicitly said the Mask is "appropriate" for the town and the council "won't be for turning" and, oddly, he also alleged (without evidence) that people had been forced into signing the petition. Alongside this hard stance, the councillors attempted to placate the campaigners with vague suggestions of a second statue. However, the campaign group rejected this with Donnelly giving the analogy that the fallacious erection of a Union Jack to represent one part of Athlone would not be undone by erection of an Irish tricolour in another. In the end, no solution to the impasse was found.

---

[14] "'Mask of the Shannon' sculpture garnering much media attention", *Athlone Advertiser*, 22 August 2019

[15] "'Mask of the Shannon' sculpture sparks rage among section of Athlone community", *Athlone Topic*, 12 September 2019.

Kenna had been invited by the *Old Athlone Society* to give a public lecture on the same day (2 October) and he took the opportunity to invite the councillors to his talk. Unfortunately, none attended. Uniquely for the Society the talk was recorded[16] and was forwarded to the councillors. But it is unknown if any listened.

Coverage in local media continued after the "showdown meeting" and Kenna's public lecture, with the focus now shifting to the council's idea of a statue for Sinann to pacify an irate public. On 19 of October the *Westmeath Independent* reported[17] "Now it appears that both of these sculptures might come to pass — with one potentially being located on either side of the river in Athlone." The newspaper went on to report that this possibility had been raised in another council meeting, five days after the "showdown meeting". The mayor "felt a new artwork at the castle 'should have a Shannon Goddess theme' and Cllr O'Rourke agreed." However, the newspaper also revealed that any funding for a Sinann statue would not come from the council itself but from another source. Nonetheless the council were "sure the funders would consult with [the council's] Arts Office." The same newspaper report reported on Kenna's public lecture and the stance that a statue for Sinann would not negate the offence cause by the Mask. The newspaper report quoted the councillors written statement read out at Kenna's lecture saying they "are happy with the piece as an appropriate art piece for the town."

And on it went; on 20 November there was another article in the *Westmeath Independent*[18] and again a second statue was discussed. In reference to Kenna's claims at his public lecture (2 October) that "the proposed sculpture 'claims a vacuous male god to represent our river while in Irish mythology the river is female'", the article stated: "However Westmeath County Council has maintained that the selection process for the public artwork was open and transparent." Just 19 days later, on 9 December 2019, Westmeath Co. Council was ranked "worst in country for transparency, accountability and ethics."[19] The newspaper report (of 20 November) went on to say how at its October meeting, Athlone's councillors again discussed the possibility "of installing a second artwork depicting the Irish mythological goddess, Síonnan." It ended by noting, however that this would be a process managed by the Office of Public Works and not the council.

The Corona pandemic interrupted the battle and throughout 2020, the intensity of the campaign abated. A *Freedom of Information* request submitted by Donnelly 10 months earlier finally delivered on August 2020 This showed that the commission agreement for the statue was signed on 19 August 2019, nearly three months after the campaign had got underway. The *Westmeath Independent* reported,[20] "It's also shown in the documents that the €60,000 in funding for the sculpture is coming from Westmeath County Council's own budget. The project is not part of the 'per cent for art' scheme, which sees a proportion of funds from a major public infrastructure development (in this case the Church Street revamp) set aside for public art."

---

[16] Kenna's Old Athlone Society talk is available on YouTube here: https://youtu.be/MDujKENPnQc

[17] Adrian Cusack, "'Síonann Goddess' artwork mooted for Athlone Castle area", *Westmeath Independent,* 19 October 2019.

[18] Triona Doherty, "No installation date yet for Church Street sculpture", *Westmeath Independent*, 20 November 2019.

[19] Mark Holland, "Kerry and Westmeath county councils fail transparency test", *The Irish Times*, 10 December 2019; https://www.irishtimes.com/news/politics/kerry-and-westmeath-county-councils-fail-transparency-test-1.4110062

It was obvious now that the council was intent on "laughing at the folly of the people" by instructing Breslin to carry on. The sculpture himself finally made an appearance in the pages of the *Westmeath Independent* on 24 October.[21] He was "not really sure how anybody could argue that it is based on British symbolism" and repeated what was given at the start of this section — that there were 14 such keystones on the 19th century Custom House and that they were designed by Gandon. He added "that there was an Irish parliament at this time." Athlone native Caroline Mannion easily picked these arguments apart in a letter published in the *Westmeath Independent* and repeated here in Appendix B1 (we remind also Figure 2.2).

So, there things stalemated. The council would continue to stick to their "process is process" mantra even though they are ranked worst in the country and the contract had been sealed after the protest started. They deemed the Mask "appropriate" for town even though it was seemingly universally rejected by a now informed people. They supported a second statue that they would not have to pay for despite the campaign repeatedly saying that a statue for Sinann does not negate the offending Mask; there is only one river deity for the Shannon in Irish mythology.

**I.E. The Statue at Solstice**

The statue was erected on the night of 19 December 2020 — just before Solstice "in the dark of the night, at the darkest time of year." There was no announcement, no ceremony, no celebration. It was very much a "damp squib" moment and to this day nobody has unveiled the statue (the mayor at this point was O'Rourke[22]). Even the *Westmeath Independent* did not run a report about this glum event.

And thus ends the battle of Sinann versus the Mask. The Council got their statue, but they lost the faith of the people. Their "process is process" stonewalling and vague promises of a second statue funded by money they did not have echoed their rank as "worst in country for transparency, accountability and ethics." The Mask still stands in Athlone, overgrown by foliage, unloved and unwanted.

> O, many a year upon Shannon's side
> They will sing upon moor and they'll sing upon heath
> Of the 600 people who signed with pride

---

[20] Adrian Cusack, "Contract for Athlone sculpture was signed after protests were underway", *Westmeath Independent,* 3 October 2020.
[21] Adrian Cusack, "Mast of Shannon artists defends controversial sculpture", *Westmeath Independent*, 24 October 2020.
[22] "Athlone has a new mayor", *Westmeath Independent,* 8 June 202;
https://www.westmeathindependent.ie/2020/06/08/athlone-has-a-new-mayor/

And their goddess eternal at swim beneath.

These are the closing lines of one or many poems penned about the battle it is based on an old poem "Ballad of Athlone". Both poems are published in full in Appendix B4.

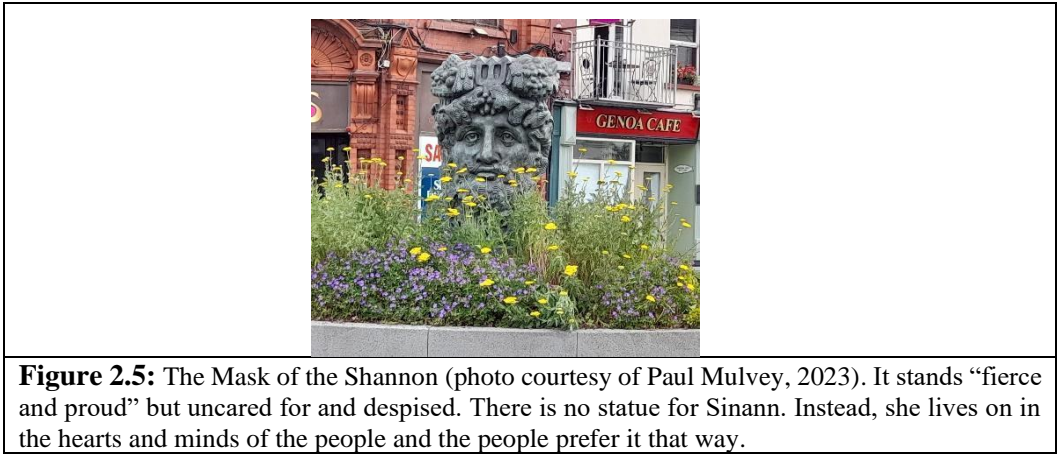

**Figure 2.5:** The Mask of the Shannon (photo courtesy of Paul Mulvey, 2023). It stands "fierce and proud" but uncared for and despised. There is no statue for Sinann. Instead, she lives on in the hearts and minds of the people and the people prefer it that way.

In the following section, we move away from the causes of the recent of the river goddess and define Sinann in her own terms. To do this we call on the people themselves to (a) do their own research into Sinann and her story and (b) to depict her artistically. We achieve both tasks through an arts project offering substantial reward to gain attention.

## II. The extraordinary story of Sinann as art

Sinann's story was far from over for, on 14 November 2020, *The Irish Post*[23] and *Westmeath Independent*[24] simultaneously[25] announced "Arts for Sinann", an arts competition with €4,000 in prizes aimed at promoting knowledge and awareness of the authentic river deity. Impressed at how a hitherto obscure goddess could inspire so many, and take on a county council, Kenna's research centre was eager to further support the MMM project and the funding was secured. So, they teamed up with *Story Archaeology*, *Rathcroghan Visitor Centre*, and *The Irish Post* newspaper to run an arts project centred on Sinann. The aim was to "reignite awareness of the mythical deity of Ireland's longest river" so that a fiasco of the type that had happened in Athlone would never happen again. In this section we observe her rise through an arts project that enabled ordinary people voice their own love for Ireland's mythological treasures.

---

[23] Fiona Audley, "Art for Sinann's Sake", *The Irish Post*, 14 November 2020.
[24] Adrian Cusack, "Shannon Goddess art competition has prizes of over €4,000", *Westmeath Independent*, 14 November 2020.
[25] Geraldine Grennan ran a parallel item in the *Westmeath Independent* titled "O'Rourke calls for 'Sionann on the Shannon!'". https://www.westmeathindependent.ie/2020/11/04/orourke-calls-for-sionann-on-the-shannon/ "We have heard a lot about Sionann over the past twelve months and I think it is noteworthy that she is not represented anywhere along the Shannon, so why not have 'Sionann on the Shannon' as part of the public realm works at the Castle" he said." Unlike the arts competition, this never came to pass.

*Story Archaeology* at the time comprised Isolde Ó Brolcháin Carmody and Chris Thompson who "are experts in Irish mythology and Sinann in particular". They commented:

> …this project springs from mathematical investigations into mythology and communicates what is so deeply appealing about both disciplines. Myths are powerful, metaphoric guides that, like tides, draw us along and Sinann is the one that flows best with this tide. The goal of the storyteller-poet, the artist, the mathematician, is to make transitions, alert and with eyes wide open. We were both delighted with this brilliant project.

Rathcroghan, in county Roscommon, is home to Cruachan Aí - the largest unexcavated Royal Site in Europe and of central importance in Irish mythology. Daniel Curley, Manager at *Rathcroghan Visitor Centre*, stated:

> As a representative of a cultural attraction which serves to interpret the rich tapestry of early Irish history, archaeology and mythology, I am always struck by the way in which our mythology captures the imagination of our visitors. Bringing these stories to the present day is enormously inspiring to enquiring minds and will ensure their survival into the future. *Rathcroghan Visitor Centre* is delighted with this endeavour and all it inspires.

*The Irish Post* "reaches a global audience through its weekly newspaper; website irishpost.co.uk and social media channels."[26] We considered it the perfect media partner to reach Ireland and its diaspora. Over the course of the competition, they ran a series of articles about various aspects of the River Shannon; industry, economics, tourism, river-life, history facts, etc. These were interspersed with reminders of the competition. The idea was to open ways for Sinann to impact the physical river. Entries were to be received by Imbolc (1 February 2021), which traditionally marks the first day of Spring in Ireland. A broad panel of 22 judges[27] (13 female, 9 male) was assembled to select the competition winners. *The Irish Post* kicked off with a brief personal interview with Kenna[28] so its readers would get to know who is behind the "Arts for Sinann project". This was followed by a similar interview with Daniel Curley[29] and then Chris Thompson.[30]

## II.A. Informing the People

The first major two-page feature[31] associated with the competition was by Editor Fiona Audley. The item was a review of a new travel guide: "to promote the Shannon as a holiday spot, giving voice to water gypsies, anglers, sailors, lock keepers, bog artists, and a water diviner celebrating wisdom through her river

---

[26] The quote comes from *Irish Post*, 6 May 2017. In the 18 May 2017 printed edition, it claimed to have had 6.51 million online users over the previous 6 months.
[27] Compare with the ``select group'' of five mentioned in the Introduction to get an idea of how representative the Council's process is.
[28] "10 minutes with Ralph Kenna", Irish Post, 8 August 2020.
[29] "10 minutes with Daniel Curley, Irish Post, 28 November 2020.
[30] "10 minutes with Chris Thompson, Irish Post, 12 December 2020.
[31] Fiona Audley, "Tales from a riverbank", *Irish Post*, 21 November 2020.

songs. Wildlife, nature and architectural as well as engineering heritage all play a part."

Next up was a feature item[32] about river pollution. Beekeeper, wildlife volunteer and anti-pollution activist Ruairí Ó Leochain explained how "in August 2019 a major oil spill had a devastating impact on the river — killing birds and fish and leaving those that survived in need of a thorough clean." Ó Leochain had since been lobbying Westmeath County Council for an Oil Response Plan but this had not yet materialised. Since *The Irish Post* item, the County Council "agreed to engage with the oiled wildlife response network". Thus, the power of the media; thus the power of mythology!

A two-page interview with Chris Thompson was hugely relevant.[33] Thompson's item outlines the importance of mythology — stories that "continue to present age-old, human problems and even solutions, that resonate, through the centuries and may still speak to us today." Thompson goes on to explain the importance of the Sinann story:

> According to one set of Dindshenchas poems, it was Sinann who brought about the creation of the River Shannon.
> The poem tells that a woman possessing great craft and skill sought to use her knowledge to help her people.
> She lacked only one skill, that of poetic inspiration.
> With this power her words could change the world. Seeking this skill, she went to Conla's realm beneath the sea.
> Here was the wonderous well, the well of Segas, or the Well of Generous Women, with all its 'lovely bubbles'.
> She won the gift she sought, but the streams that fed the well beneath the sea rose up like a great wave, covering the land, and forming the great River Shannon.
> Sinann drowned but the river brought the new and prosperous life she was seeking for her people.
>
> Like many people who go searching for a translation of the story of Sinann, I first encountered the 19th century re-telling by Eugene O'Curry.
> This version implied that she was a disobedient girl who dared to approach a secret well which rose against her impudence.
> All versions will represent the mores and concerns of the time in which they are retold.
> It is certainly true that O'Curry was a man of his time.
> It is also unfortunate that he chose to conflate the story of Sinann with that of Boann and the Boyne.
> Hers is a good story and does involve a well, but it is nothing to do with Sinann.
> There is no hint of disobedience or impudence in the Dindshenchas poem of Sinann. She is a poet of skill who, by right, may seek to enter the Otherworld realm beneath the waves.

---

[32] Fiona Audley, "Saving the Shannon", *Irish Post*, 5 December 2020.
[33] Chris Thompson, "Celebrating Sinann", *Irish Post*, 26 December 2020; https://www.irishpost.com/life-style/why-the-story-of-the-goddess-of-the-river-shannon-is-one-worth-telling-200145

Through this item in *The Irish Post* and the *Story Archaeology* website, Sinann's story was available to the people. Moreover, it corrected a flawed Victorian account of a "disobedient girl" by one Eugene O'Curry. Artists thus had the correct version to hand. Indeed, diligent researchers had only to go to the *Story Archaeology* website for Ó Brolcháin Carmody's translation and Thompson's interpretation. We reproduce both of these in the next section.

**II.B. Arts for Sinann**

The project resulted in 47 high-quality submissions being received (63% of artists were female) from at least 7 countries including Ireland (17), England (12), USA (5), Spain (3), Tasmania (1), Wales (1) and New Zealand. Submissions came in a diversity of forms including: paintings (16); Poems (11); digital art (5); community art performances (3) ; etching (3); short stories (2); photos (2); enamel and skillet pots (2); dance (1); movies (1); drawings (1); embroidery (1); a "lesson" (1); a 192- page e-Book (1), a play (1), stained glass (1); a spirit doll (1); a collaborative work that weaves together painting and poetry in both English and Irish.

An algorithm called *Calibrate with Confidence* was used to adjust for different levels of stringencies and confidences in the 22 judges. Invented by Robert MacKay, this algorithm had been developed with Kenna a few years earlier [MacKay et al,2017]. It is superior to the commonly used approach in panel assessments where different submissions are assigned to different assessors and simple averages are taken. Any assessment protocol would not expect each judge to assess 47 objects (only about 15 items were assigned to any individual in "Arts for Sinann") and using simple averages opens a risk of some entries being assigned to overly stringent or generous reviewers. The algorithm takes this form of Lady Luck out of the equation by making sure the network of assessed objects is intact so that every assessor can be calibrated against every other. It also takes different levels of confidences in judges scores into account. Demonstrating the impact of this algorithm was another aim of this competition and, indeed, it did make a difference; one of the winning pieces had been assessed by stringent judges and would not have been successful had simple averages been taken. It turned out that reviewers with humanities backgrounds tended to be most stringent in this process with lay reviewers most generous, and scientists are in the middle. The winning piece had been assessed by five humanities judges, two lay and only one scientist. These differences in disciplines are not something we anticipated and, again, demonstrated the ongoing research nature of the project.

The results of the Sinann competition were announced in a two-page spread in *The Irish Post* on 27 March 2021 with the triumphant headline "Art competition awakens knowledge of forgotten Irish goddess."[34] The *Westmeath Independent* followed suit with another two-page feature to announce the results.[35] The three winning pieces are displayed in Figure 3.1. These are also depicted, alongside the full contextual information given by the artist in the *Story Archaeology* website[36]. The website includes a gallery with many of the other submissions. The following week *The Irish Post* ran another two-page spread showcasing some of the remaining entries. Children's entries featured on 15 May with every entrant receiving a prize. Indeed, every adult entrant and assessor also received a small gift from *Rathcroghan Visitor Centre* along with constructive feedback on the quality of their entrance. This positivity was to demonstrate a shared goal of promoting Sinann rather than competing against each other.

---

[34] Fiona Audley, "Art competition awakens knowledge of forgotten Irish goddess", *The Irish Post,* 27 March 2021.
[35] "Locals among winners of art for Sinann competition", *Westmeath Independent*, 8 May 2021.

| 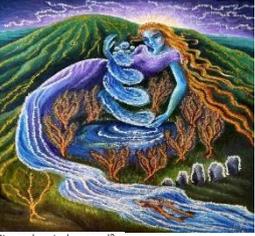 | 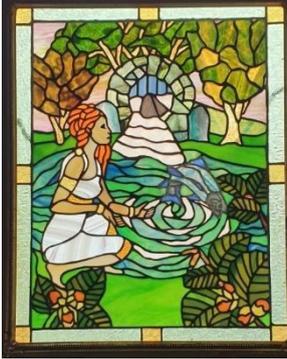 | 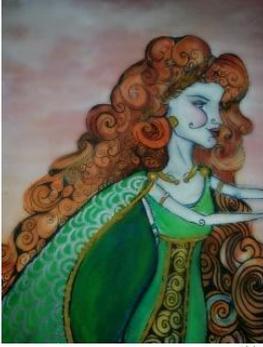 |
|---|---|---|
| **"Sinann"** | **"Sinann in Stained Glass"** | **"An Sinnann"** |
| A collaborative work that weaves together painting and poetry. Drawing inspiration from the three Dindshenchas stories about Sinann, our work focuses on Sinann's journey seeking imbas forosnai, that moment of transformative illumination as she meets the waters of the well, and her entry into her "new life"../ | Depicts Sinann at Connla's Well with the rising waters of the Shannon Surrounded by the Hazel trees who's fruit when eaten gave knowledge to the salmon. Height X width - excluding the wood frame: 20 X 16 inch. Materials: Stained glass, copper foil, 40/60 lead solder. Pieces: 486 pieces. | Acrylic painting inspired by the modern retelling of the An Sinnann story as seen ó. storyarchaeology.com. 12"x32" inch pre-stretched deep wedged canvas Portrays Sinann as a strong and beautiful woman much like the river who adds such beauty and grace… |
| Morpheus Ravenna and Caróg Liath | Lee Fenlon | Avril Egan |

**Figure 3.1:** Winning submissions to the Arts for Sinann competition.

Alongside demonstrating that Ireland could produce art of sufficient high quality for local and, indeed, national, iconography, a purpose of "Arts for Sinann project" was to demonstrate what Sinann means to people. Many unsolicited positive and grateful remarks were received and here we document some of those remarks:

- Even before I heard of the competition, I had wanted to depict Sinann as a female form after the mask fiasco in Athlone.
- Please pass on my appreciation of comments made on each of the submissions. It was a joy to do…You have done a fine job highlighting Sinann and I hope that it will take on momentum now. I believe it will.
- I was thrilled to hear the judges' comments, I had never sent a poem to a competition before.
- I really enjoyed getting to know all about the story of Sinann
- I have never in my life entered an art competition! This subject just felt very close to my heart. It is exceptionally good of you to give me the critical remarks, I appreciate it...I am happy to have helped raise the profile of this beautiful legend.

---

[36] https://storyarchaeology.com

- I very much enjoyed working on this piece and hope I will take part in future events. Thank you again for the opportunity…I'm not sure I would want to compete, it's just that the story needed an illustration, to me Sinann needs a human form, a poetic embrace, a story, told in the old ways. Thank you for the chance to channel my memories into a work of art.
- Sinann is very real to me and I wanted to capture the truth of her essence in this story.
- You're taking me well out of my comfort zone which is exactly what I need right now, so my thanks to you!
- Thank you so much for the pleasure of having this creative project over this winter. And thank you for honoring this special goddess and river. It brought great joy to my heart to celebrate along with all who are participating.
- I have not written anything like this before, and nothing much since my dissertation on Sovereignty. It was wonderful to have the opportunity to offer tribute to Sinann and since writing it, I have deepened even more into this living myth inside of me, She continues to offer healing…It was never about winning, but rather about making the offering, the ritual of fostering connection. There isn't a day goes by when I don't long to return to Ireland.  [an artist from USA]
- Perhaps I will pull inspiration from our correspondence to update my work on Sovereignty and turn it into a proper book. Please do advise where I should continue to look for future competitions. Eimear sent me the Story Archaeologists non-patriarchal version of Sionnan's story and that is how I came to you.
- When I found out about the competition, I was going through a period of uncertainty and doubts. Even though I knew that I wanted to try something different in my creative practice, I wasn't confident enough to start doing something new. Arts for Sinann gave me a safe space and motivation to experiment. I'm very grateful for having had this opportunity.
- I am reading one of your papers now because of finding this contest and also saw you are on Facebook and sent a request to connect to be aware of your ongoing research. I love the threads of connection that come together in this contest and in your ongoing collaborations and work.
- While I have you, is there any site you can direct me to that can tell me the story of Boinne? I have been asked to create an illustration around her, but what I found is very similar to Sinann's back story!

Thus, we end Section II having evidenced what Sinann means to people and having shown that Ireland has sufficient talent to depict her accurately and attractively.

In Section III we move on to give Isolde Ó Brolcháin Carmody's new translation of Sinann's story from the Metrical Dindshenchas and Chris Thompson's interpretation of same.

## III. The Extraordinary Story of Sinann – The Authentic Translation

On the 18 June 1857, Eugene O'Curry delivered a lecture entitled "Of Education and Literature in Ancient Erinn", which was later transcribed and published by William Sullivan [O'Curry, 1873]. It was at this event O'Curry offered a now famous version of the story of Sinann. We reproduce it for completeness in Appendix C with accompanying notes by Isolde Ó Brolcháin Carmody. His version told of a young girl who had dared to approach a well, hidden as it was among nine hazel trees whose fruit fed the salmon of wisdom. This well, he added, was guarded by the wisest of men. Her approach caused the well to rise up, overflow its bounds, and to become a river. Her lifeless body was carried away in the flood which eventually became the Shannon. Thus, the river is named after her.

**III.A. Instinct: Sinann is More than a "Disobedient Girl"**

When first encountering this story, Thompson found its theme somewhat troubling. As a mythologist, she was aware of Boann and the creation of the river Boyne as taking a very similar form.[37] Folklore versions also featured floods after a woman approached, or removed, a cover from, a well. As a storyteller, Thompson placed emphasis on the life-giving creativity of the river's release and her instinct suggested similar might be at play in Sinann's case. This instinct turned out to be quite correct, as we shall see.

Although her lineage is mentioned elsewhere, the story of Sinann is found in *Acallam na Senórach (Colloquy of Ancients),* a textual compilation only found in the Metrical Dindshenchas and thought to have been put to writing in the thirteenth or fourteenth century. The Dindshenchas, or *Lore of Placenames,* is a class of onomastic texts forming a collection of early oral material curated in, or around, the 12[th] century. The first recension of the Dindshenchas in the *Book of Leinster* contains the story of Sinann.

In the early part of the twentieth century, Edward Gwynn compiled and translated Dindshenchas poems from *Lebor na hUidre*, the *Book of Leinster*, the *Rennes Manuscript*, the *Book of Ballymote*, the *Great Book of Lecan* and the *Yellow Book of Lecan.* He published these in four parts between 1903 and 1924 with a general introduction and indices published as a fifth part in 1935 [Gwynn, 1906]. Eugene O'Curry, therefore, did not have this essential resource to draw on for his 19[th]-century translation of Sinann's story.

Although the date of the written compilation of the material is clearly post Norman, some of the Dindshenchas concerns older local genealogy, connecting people to places, and there is also earlier lore that remembers information no longer extant. This includes placenames that would have fallen out of use for hundreds of years. Another function of the Dindshenchas poems is to record and explain the origin of landscape features. For example, one poem relates the creation of the river Barrow as having been created by the meanderings of a great péiste.[38] The Dindshenchas story of Sinann belongs to this category. Earlier dating of material may also be supported by the way several alternative versions of Dindshenchas poetry being attached to the same place.

---

[37] From the Metrical Dindshenchas, Volume 3, poem 3pp 34 – 39 [Gwynn, 1906]. The Dindshenchas relates how Boann gave birth to Oengus, the 'young son', so called because the sun was stopped in the sky to allow the time from conception to birth to take place within one day. In this Dindshenchas, Bóand seeks to bathe in a guarded well to disguise her 'adultery' and the water rises up and drowns her. This story is thematically closer to O'Curry's version of Sinann. It is interesting that the Dindshenchas version of the story of Bóand is the one that is generally known, whereas the O'Curry story of Sinann largely replaced the older version.

[38] From Volume 2 of [Gwynn, 1906] poem 13, p. 62. A version translated by Isolde Ó Brolcháin Carmody, based on Gwynn's work is available on the *Story Archaeology* website.

In the very first episode of the *Story Archaeology* podcasts, dated June 2012, Thompson, and Ó Brolcháin Carmody explored the story of Sinann using Gwynn's compilation of the Dindshenchas. They revisited and updated their research in August 2015. It was in that examination of the Dindshenchas that they came to understand just how deep were the deep the discrepancies were between this version of Sinann's story and the better-known 19th c version, as given by O'Curry.

The Dindshenchas tale of Sinann, as found in Gwynn's Metrical Dindshenchas (Vol 3, poem 53, p. 286 ff of [Gwynn, 1906]), tells of a young and gifted young woman. She is described as noble, compassionate, radiant, powerful and of high repute. She goes on a quest to seek "inspiration". This suggests that she is a poet who seeks poetic inspiration, probably *Imbas forosnai*, a form of visionary poetry, to make a new land for her people. Nowhere does the poem suggest that she is disobedient or that she has gone beyond her expected rights.

The other main difference between the new interpretation and that of O'Curry is that the well she seeks is in Connla's realm[39] '*under the blue rimmed ocean*'. In one version she goes to seek the Well of Generous Women, which is given that name to celebrate her name. Her quest, which costs her mortal life, is regarded as a gift to her people. Unfortunately, she is not regarded with the same respect by O'Curry.

Sinann makes one other appearance in the two Dindshenchas poem of Áth Líac Find (the Ford of Finn's stone).[40] Both poems recognise this ford as a site of a significant battle where Finn was victorious. The battle is described in some detail; terms such as the defeat of 'fifties' symbolically indicate involvement of very large numbers. Names of defeated heroes are also given as multiples, for example, there are four Conalls, two Colmans and other multiple names. Attention is deliberately given to named individuals or groups of defeated heroes, such as the three sons of Cirb.

After the battle, Sinann, here described as the daughter of Mongán[41], presents Finn with a stone on a golden chain. This is thrown into the river where it will stay until Sinann, now given the title of Bé Thuinne, Lady of the Wave, will retrieve it one Sunday morning. Seven years after that, so says the poem, will come the end of the world. This appearance, once more, defines Sinann in the role of the poet herald (fili) recording the victory and proclaiming its projected future consequences. All of this presents Sinann as a powerful fili (poet herald) and strengthens her connection with the original Dindshenchas poem. The connection between Sinann and the Boand dindshenchas becomes quite a gulf!

---

[39] The reference to Connla's realm supports the location of this well in the Otherworld beneath the sea. The middle Irish literary tale, *Echtrae Tadhg Mac Cein* (The Adventures of Tadhg Mac Céin) also tells how the Otherworld realm was still associated with Connla, who greets Tadhg in this later but highly entertaining tale.

[40] From Volume 4 of [Gwynn, 1906] Poems 11 & 12 pp 36-43. A version translated by Isolde Ó Brolcháin Carmody, based on Gwynn's work is available on the *Story Archaeology* website.

[41] Mongán is an important character. His birth as a future fili and king is prophesied in an early text "Immrám Brain Mac Febul (The Voyage of Bran). He remained a significant character, surviving in several medieval tales. In "The Colloquy of Colmcille and the Youth at Carn Eolairg", a fragmentary poetic text, he appears as a mysterious youth and communicates secret ancient dindshenchas knowledge to Colmcille.

So why did Eugene O'Curry not pass on her story in the form it appears in the original Dindshenchas? Any answer must involve speculation, of course and there is several possibilities.

Eugene O'Curry, *Eoghan Ó Comhraí*, was a highly reputable scholar, who undertook significant and important work in the field of Irish studies. In 1835 O'Curry was employed by the ordnance survey and worked closely with John O'Donovan in gathering material concerning ancient sites as well as recording information from knowledgeable local informants on place names and family names. His first language was Irish, and he worked closely providing translations and notes for the Celtic Society, founded in 1845.

However. he was working at an early date without the benefits of modern scholarship including the current onomasticon. He was also a man of his time. In the mid 19$^{th}$ century women generally had lower status in society and were not permitted to take prominent roles. He may have viewed Sinann's story through this lens.

He may also have assumed that the stories of Boann and Sinann were two versions of the same story. He may have felt that, without including context relating to the role of the poet (fili) in early Irish society, the Dindshenchas poem would have appeared problematical or obscure. O'Curry was indeed keen to popularise the Irish textual stories and conflating the Sinann story with that of Boann may have made it easier to follow for an audience who no longer encountered these stories on a regular basis.

There is one more strand to the journey the story has taken. In 2014/2015 A reading of the Dindshenchas story of Sinann implied that when Sinann gained her quest, and the power of inspired poetry, the lovely bubbles rose up in a wave and covered the land carrying the salmon of wisdom and forming the new river further inland. Aware that this was no more than pure speculation, we did propose the possibility that this might have represented some kind of major flood event, on the west coast of Ireland, sweeping away the coastal Macha flatlands and leading to a move inland. We refer the reader to the *Story Archaeology* website to pursue this further.

**III.B. The New Translation of Sinann's Story**

On 21 August 2015, Isolde Ó Brolcháin Carmody published a new translation of the extraordinary story of Sinann.[42] This translation has never been published in print in a peer-reviewed academic journal before. Given its importance we reprint it here in full.[43]

---

[42] From translated by Isolde Ó Brolcháin Carmody's translation of the Metrical Dindshenchas, Volume 3 pp. 286–297; poems 53 and 54 [Gwynn, 1906].
[43] Reprinted from: https://storyarchaeology.com/the-poems-of-sinann-2

|   | The Poems of Sinann as posted on *Story Archaeology* on 23/06/2012, translated by Isolde Ó Brolcháin Carmody from the Metrical Dindshenchas, Vol. 3 pp. 286–297; poems 53 and 54 [Gwynn, 1906] | |
|---|---|---|
|   | **Sinann I** | |
| 1 | Sáer-ainm Sinna saigid dún, | Seek Sinann's noble name for us |
|   | dáig rolaimid a lom-thúr: | as you strive to uncover its origin: |
|   | nirb imfhann a gním 's a gleó | The deeds and the struggle were not insignificant |
|   | dia mbói Sinann co slán-beó. | That made Sinann whole-alive [i.e. immortal?]. |
| 2 | Rop ingen rogasta ríam | There was once a very powerful girl – |
|   | Sinann sholasta shír-fhíal, | Sinann, radiant and ever-compassionate |
|   | co fúair cach ndodáil nduthain | Until she met an all-encompassing death |
|   | ingen Lodáin laech-luchair. | That daughter of Lodán from heroic Luachair |
| 3 | Hi tír tarngire co túi, | In the restful Land of Promise, |
|   | ná geib anbthine imchrúi, | Unsullied by storms of blood, |
|   | fúair in suthain-blaid rosmill | The eternally reknowned [one] met destruction |
|   | ingen Luchair-glain lúaidimm. | That girl from pure Luachar whom I laud. |
| 4 | Tipra nad meirb fon muir mass | There was a never-stagnant well beneath the pleasant sea |
|   | for seilb Chondlai, ba comdass, | in Connla's realm, appropriately, |
|   | feib adrímem ria rélad, | As we enumerate in the telling of the tale, |
|   | luid Sinann dia sír-fhégad. | Sinann went to gaze upon it eternally. |
| 5 | Topur co mbara búaine | The well perpetually flows |
|   | ar ur aba indúaire, | on the brink of a bright, cold river, |
|   | feib arsluinnet a clotha, | As its renown is retold, |
|   | asmbruinnet secht prím-shrotha. | Seven great streams brim forth [from it] |
| 6 | Immas na Segsa so dait | Inspiration of Segas is found here |
|   | co febsa fond fhír-thiprait: | excellently, under the true-well: |
|   | ós topur na tond tréorach | Before the well of the strong waves |
|   | fail coll n-écsi n-ilcheólach. | Stand the many-musicked hazels of the scholars. |
| 7 | Síltair sopur na Segsa | The foam of Segas is sown |
|   | for topur na trén-chennsa, | over the well of the strong-gentle one [fem.] |
|   | ó thuitit cnói Crínmoind cain | Where Crínmond's sweet nuts fall |
|   | fora ríg-broind réil roglain. | Onto its bright, very pure regal breast. |
| 8 | In óen-fhecht n-a tuile thrumm | All at once, in a heavy flood, |
|   | turchat uile don chóem-chrund, | All erupt from the shapely tree, |
|   | duille ocus bláth ocus mess, | leaf and flower and fruit |
|   | do chách uile ní hamdess. | For everyone – it is not unlovely! |
| 9 | Is amlaid-sin, cen góe nglé, | It is like this, clear without falsehood, |
|   | tuitit n-a róe dorise | They then fall in their season |
|   | for topur sográid Segsa | Onto the beloved well of Segas |
|   | fo chomdáil, fo chomfhebsa. | In the same moment, with the same excellence. |
| 10 | Tecait co húais, ra gním nglé, | They come nobly, with clear action, |
|   | secht srotha, búais cen búaidre, | the seven streams, their gushing unhindered, |
|   | dorís isin topur the | Again into that well |
|   | dianid cocur ceól-éicse. | Causing whispers of musical knowledge. |
| 11 | Adrímem in uide n-úag | We shall recount the whole journey |
|   | dia luid Sinann co sóer-lúad | whereon went Sinann of noble repute |
|   | co lind mná Féile fuinid | to the Pool of the Generous Woman where the sun sets [i.e. in the West] |
|   | cona gléire glan-foruid. | With the best of her pure household. |
| 12 | Ní thesta máin bad maith linn | There is no lack of any desirable talent, |
|   | for in saír sin ná saílfinn, | In that noblewoman, that I could imagine, |
|   | acht immas sóis co srethaib, | except for streams of expert inspiration |
|   | ba gním nóis dia núa-bethaid. | It was a new activity for her new life. |
| 13 | Rotheich in topur, toirm nglé, | The well turned back, a clear sound, |
|    | tria chocur na ceól-éicse,               | Through the whisper of the musical knowledge,                |
|    | re Sinainn, rothadaill túaid,            | before Sinann, who touched it in the north,                  |
|    | cor-riacht in n-abainn n-indúair.        | Until she reached the truly-cold river.                      |
| 14 | Rolen sruthair na Segsa                  | The stream of Segas followed                                 |
|    | ben Luchair na lán-gensa                 | the fully virginal woman of Luachar                          |
|    | cor-riacht huru na haba                  | Until she reached the brink of the river                     |
|    | co fúair mudu is mór-mada.               | So that she met destruction and great frustration.           |
| 15 | Andsin robáided in breiss,               | Then the beauty was drowned,                                 |
|    | is rothráiged fo throm-greiss:           | and she ebbed [sic!] under heavy blows;                      |
|    | cid marb in ben co mbruth baidb          | Although the woman of bubbling crow[-energy] is dead,        |
|    | rolen dia sruth a sáer-ainm. S.          | Her noble name follows her river.   S[inann].                |
|    | Desin fri déine ndile                    | From then, with swift affection,                             |
|    | lind mná Féile fír-gile:                 | is [known / called] the truly-shining Pool of the Generous Woman: |
|    | fail cech óen-airm, cúairt n-assa,       | in every single place, an easy journey,                      |
|    | sáer-ainm súairc na Sinna-sa. S.         | the pleasant, noble name of this Sinann [is known].   S[inann.] |
|    |                                          |                                                              |
|    | **Sinann II**                            |                                                              |
| 1  | Sinann, cá hadbar diatá,                 | Sinann, what is the cause of it [its name],                  |
|    | inneósad cen immargá:                    | I shall tell without deceit:                                 |
|    | atbér cen snaidm co solus                | I will say [it] clearly without complication                 |
|    | a hainm is a bunadus.                    | Its name and its foundation.                                 |
| 2  | Innisfed do chách uile                   | I will tell to everyone                                      |
|    | bunad Sinna srib-glaine:                 | the origin of pure-streaming Sinann:                         |
|    | ní chél in dag-blad diatá:               | I will not conceal the reason of its good repute,            |
|    | atbér adbar a hanma.                     | I will say the cause of its name.                            |
| 3  | Tipra Chonnlai, ba mór muirn,            | Connla's well, it was clamourous,                            |
|    | bói fon aibeis eochar-guirm:             | It was under the blue-rimmed ocean:                          |
|    | sé srotha, nárb inann blad,              | Six streams, of unequal fame,                                |
|    | eisti, Sinann in sechtmad.               | rose from it, the seventh was Sinann.                        |
| 4  | Nói cuill Chrimaill, ind fhir glic,      | The nine hazels of Crimall, the clever man,                  |
|    | dochuiret tall fon tiprait:              | cast [fruit] onto that well:                                 |
|    | atát le doilbi smachta                   | They are [there] under mysterious control                    |
|    | fo cheó doirchi dráidechta.              | Under a gloomy fog of magic.                                 |
| 5  | I n-óen-fhecht, amail nách gnáth,        | In the same moment, in an uncanny way,                       |
|    | fhásas a nduille 's a mbláth:            | their leaves and their flowers grow:                         |
|    | ingnad ciarsad sóer-búaid sin            | A wonder is this, though a noble trait,                      |
|    | 's a mbeith i n-óen-úair abaig.          | As is their ripening at the same time.                       |
| 6  | In úair is abaig in cnúas                | When the cluster of nuts is ripe                             |
|    | tuitit 'sin tiprait anúas:               | they fall down into the well:                                |
|    | thís immarlethat ar lár,                 | they scatter below on the surface,                           |
|    | co nosethat na bratán.                   | So that the salmon eat them.                                 |
| 7  | Do shúg na cnó, ní dáil diss,            | From the juice of the nuts, no insignificant matter,         |
|    | dogníat na bolca immaiss;                | are formed the bubbles of inspiration;                       |
|    | tecait anall cach úaire                  | In this way, they come every hour / time                     |
|    | dar na srothaib srib-úaine.              | On the green-flowing streams.                                |
| 8  | Bói ingen, ba buide barr,                | There was a girl, her hair was yellow,                       |
|    | thall a túathaib dé Danann,              | There from the Tuatha De Danann,                             |
|    | Sinann gasta co ngné glain               | Powerful Sinann with her pure face,                          |
|    | ingen Lodain luchair-glain.              | daughter of Lodan, from pure Luachair.                       |

| 9  | Smuainis ind ingen adaig, | One night, the girl had an idea, |
|    | in bind bél-derg banamail, | that melodious, red-lipped woman, |
|    | co mbói da hindus cach mblad, | With all reknown in her gift, |
|    | acht in t-immus a óenar. | Except only for inspiration. |
| 10 | Lá da tánic cosin sruth | The day that she came to the stream |
|    | ind ingen, ba cóem a cruth, | that girl, her form was shapely, |
|    | co facca, nochor dál diss, | so that she saw, no insignificant matter, |
|    | na bolca áilli immaiss. | The beautiful bubbles of inspiration. |
| 11 | Téit ind ingen, toisc úaille, | The girl goes, lamentable quest, |
|    | 'na ndiaid 'sin sruth srib-úaine: | after them into the green-flowing stream: |
|    | báiter hí da toisc anall; | she is drowned there through her quest; |
|    | conid úaidi atá Sinann. S. | so that from her is Sinann [named].   Sinann |
| 12 | Dénum aile, mad áil lib, | Another version, if you wish, |
|    | uáim ar in Sinainn srib-gil, | from me, about the bright-flowing Sinann, |
|    | cé bethir lim 'ca légud, | though it is to be read in my verse, |
|    | ní ferr hé 'ná in cét-dénum. | it is not better than the first version. |
| 13 | Lind mná féile, ba fír dam, | The Pool of the Generous Woman, I have it true, |
|    | ainm na linde 'nar 'báided: | is the name of the pool where she was drowned: |
|    | is é a dír maras dise, | It is fittingly [named] from her |
|    | más fír é fri indise. | If it is true to tell. |
| 14 | Dénum aile, is mebair lemm, | Another version I have in mind, |
|    | rochúala cách co coitchenn; | everyone in general has heard: |
|    | Cú Núadat, ba mór maise, | The Hound of Núada, great was his beauty, |
|    | robáite 'sin chrúad-glaise. | was drowned in that cruel stream. |
| 15 | Nó combad Sinann co becht | Or perhaps Sinann is literally |
|    | Sín Morainn, tre etercherт: | Morainn's Collar, by interpretation: |
|    | nó sí in moirenn, aidble gním: | or "she is the sea", might of deeds: |
|    | áille Sinann 'ná cach sín. | Sinann is more beautiful than any storm. |
|    |  |  |

Accompanying Ó Brolcháin Carmody's translation are a set of notes on name and places which we reproduce in Appendix C1. We also re-publish O'Curry's telling of the story of Sinann [O'Curry,1873] with accompanying critical notes by Ó Brolcháin Carmody in Appendix C2.

The material in this section is presented with the hope that future descriptions of Sinann are accurate and not drawn from a flawed Victorian translation. Next, we turn the wider legacy of that era and the wider issue of colonialism. The sorry saga of Athlone's statue illustrates that it has not at all gone away but is deeply endemic in society. Its close friend misogyny accompanies it.

## IV. The Origins of Colonialism and its Legacy in Ireland Today

The ubiquitous presence of neoclassical iconography of Empire statuary (the stone 'god' heads on the parapet of the Custom House in Dublin being but one example) demonstrates the persistence of colonial signs, signals and insignia with a long lineage that reaches back to the initiation of the English colonial campaign in Ireland. Other prominent examples are the Wellington Monument in the Phoenix Park celebrating British Imperial wars, or the statue of Prince

Albert still on display in the grounds of Leinster House, and royalist nomenclature such as the *Royal* Dublin Society, the *Royal* Academy of Music, the *Royal* Irish Academy, etc.

It was the distinctly anti-Irish Giraldus Cambrensis, in his *Topographia Hibernia* of 1188, who set the scene for the 'othering' of the Gaelic Irish—an essential agenda for what would become the longest colonial project in history. The sheer length of the English/British colonial project in Ireland — arch-colonialist T.B. Macaulay [Pine, 2014] considered it 'the greatest colony that England had ever planted' — applied an overwhelming degree of pressure on Gaelic mythology, identity, and its socio-political structures.

A point repeatedly made by apologists for English colonialism is that the indigenous in Ireland had a tenuous hold on a consciousness of nationhood, and thus colonialism did not have such a debilitating impact on a nation that was, in any case, 'questionable', that the entire colonial project could be justified as a 'civilizing' mission for a 'barbarian' people not yet fully formed into a cultivated nation proper. The fact is, however, that there was real integrity to the way the Irish understood themselves to be; there were well-formed elements of nationhood before the English invasion (or *expugnatio*).[44] This is a point that is insufficiently argued and thus little understood. Except for those from the most retrograde quarters, it is now unchallenged that the Irish-Gaelic language is Europe's earliest surviving vernacular literary tradition. And while the Irish did not emulate England's political model of 'state' (why should they?), the Gaelic Irish were very conscious of their ethnicity, their distinct mythological, oral and written traditions. As Adrian Hastings remarks (in comparing Gaelic Wales and Ireland) [Hastings, 1997]:

> Early medieval government and political identity were quite as much a matter of law and custom as of anything that a named ruler might be or do, and both Wales and Ireland had developed an extensive vernacular law recognised as authoritative throughout the land. Wales was politically one to the very real extent that it had one law, the law of Hywel Dda, and Ireland was similarly unified by Brehon law […] Language, law, literature, a sense of historic identity, a particular kind of culture sustained by the orders of bards, jurists and monks, that surely is sufficient to show that…Ireland [was] by the mid-eleventh century well past the dividing line between ethnicity and nation.

However, the keystones of early Irish 'nationhood'— mythology, land, language, culture and religion — having all come under severe and prolonged colonial assault following centuries under the increasing domination of English colonization, meant that deep traumas were inflicted on every aspect of Gaelic identity.

---

[44] Thomas Bartlett's reading of Ireland's colonial history makes it clear that 'what happened was an invasion, followed by a conquest of a large portion of the island; all attempts to portray the invaders as if they were guests of an Irish king, or medieval tourists…fail to recognize the determination of the invaders [who] proudly described their action as a conquest (*expugnatio*) for centuries thereafter…It was an English invasion…: all talk of the "Normans", or the "Anglo-Normans", or "Anglo-French", or even the "Cambro-Normans" coming to Ireland is simply ahistorical. The invaders called themselves English (Engleis, Angli), were called Saxain (= English) or *Gaill* (= foreigner) by the Irish, and for the next seven hundred years were designated as English in the historical literature. Contemporaries never described them as Norman, Anglo-Norman…Only in the late nineteenth century, and largely on grounds of political sensitivity, was the identity of the English invaders fudged by these non-historical terms.' See [Bartlett, 2010].

While land dispossession started immediately with the arrival of the English in the 12th century, the Tudor and (later) Cromwellian conquests saw an immense transfer of land from Catholics to Protestants. The cultural and political restructuring of Ireland reached its zenith in the era that spanned the period of the Tudor and Stuart monarchs. With the Flight of the Earls in 1607, the final symbolic remnants of the old Gaelic civilization had been largely dismantled and ultimately replaced by the institutions of Protestant Ascendancy dominance. This dominance was further enhanced by Cromwell's 1642 Adventurers Act whereby financial investments in support of the conquest of Ireland were rewarded with land: 'Six shillings was the price of an acre in Connacht, while ten shillings and twelve shillings were the respective values of an acre of land in Munster and Leinster.' For just one shilling, Adventurers received an acre of Ulster land (where resistance to English rule was at its highest pitch).

Anthony Cronin correctly points out that because many in Ireland 'believed that huge acreages had been stolen; folk tradition and religion ensured that this historical theft should, unlike other such grand larcenies, not be allowed to become part of the normal order of things.' [Cronin, 2002] The extent to which people's actual lives and ways of life were violated by such land appropriation should not be underestimated. Many of the occupied territories had strongly felt sacred, tribal and mythological associations that had been deeply rooted for centuries, if not millennia. The Cromwellian cheap sell-off of land and its attendant ethnic cleansing — 80% of acreage was in English hands by 1700 — was to have disastrous consequences that resonated well into the future, ultimately partitioning the island with devastating results.

Consequently, the intellectual, poetic, legal, socio-cultural and mythological elements of Gaelic society and Gaelic collective memory dissipated in the wake of the old order's collapse. 'After the seventeenth century the assurance of this discourse waned…the native institutions which had hitherto supported Gaelic poetry — the whole educational, legal, religious and economic continuum — virtually disappeared.' As an oral tradition that was less and less practised, the richest and most elaborate expressions of Gaelic music, poetry and mythology slowly and steadily declined as a thriving cultural expression. The virtual extinction of Ireland's rich Gaelic civilization thus serves as both a poignant and ironic subtext for the use of the medieval Irish harp today as the country's national emblem. As the Gaelic arena disappeared, it was replaced with the fiscal and political structures, and the cultural and iconographic narratives of the coloniser.

**IV.A. The Colonial Mindset and its Impact on Ireland**

Building on what we know of Cambrensis' racist characterizations of the Gaelic Irish, and mindful what T.B. Macaulay's statement above, let us look a little more closely at the mindset of the colonialist, particularly in relation to his prerequisite to 'other' the colonised, to destroy its culture and replace it with so-called 'civilised' constructions of identity that pervade all elements of socio-cultural and political arenas.

Edward Said tells us of the relationship between Ireland and its long-time colonial ruler, how sectarianism is a central agent of colonialism, and how it establishes what he calls elsewhere the native's 'civilizational inferiority' [Said, 1993]:

> The high age of imperialism is said to have begun in the late 1870s, but in English-speaking realms it began well over seven hundred years before…From that time on an amazingly persistent cultural attitude existed toward Ireland as a place whose inhabitants were a barbarian and degenerate race.

To value Irish ethnic principles, customs and mythology would be to value the cultural traditions that supremacist forces had spent generations successfully destroying. Thus, central to the logical mindset of the colonizer lies an imperative to suppress any form of expression that might strengthen the cultural identity of the colonized. As Franz Fanon argues [Fanon, 1968]:

> Colonialism is not satisfied merely with holding a people in its grip and emptying the native's brain of all form of content. By a kind of perverted logic, it turns to the past of the people, and distorts, disfigures and destroys it.

In Ireland, therefore, the Gaelic language, Gaelic customs, Gaelic mythologies, Brehon laws, Gaelic social structures and Catholic affiliation were all continually and successfully undermined through ongoing colonial plantation, disenfranchisement, confiscation and the Penal Law suppressions. This project coincided with an ongoing endeavour towards anglicization. As cultural forces, Gaelic mythology, language, and culture (arts) were considered by the English as major obstacles to their colonizing project. This fear instigated a mission of cultural genocide in Ireland of which the anglicization of the country played a central role. As Attorney General for Ireland from 1606, Sir John Davies stated: 'We may conceive and hope that the next generation will in tongue and heart, and everyway else, become English, so as there will be no difference or distinction but Irish Sea betwixt us.' [Kane, 2011] The reason why Gaelic culture and mythology were seen as a danger to colonial rule was that they acted as repositories for Ireland's past and its history. The erasure of Gaelic cultural and mythic memory was thus central to the success of the colonizing mission. As Edmund Spenser proposed in a tone of svelte enunciation of censor and slaughter, '[Gaelic] books and antiquities…should be seized, and carefully translated, into English and then burnt, and the professors very carefully dealt with.' [Spenser, 1997]

As mentioned, from the 12th century, but increasingly from the 1600s, Ireland's population was subject to endless waves of land dispossession, religious intolerance, and deracination from the most fundamental cultural elements of language, poetry, music and mythology. In many respects, a natural result of this ongoing subjugation was the so-called 'Famine' (1845-1848). Under British administerial governance and fuelled by its laissez-faire economic policies (signifying that catastrophe could always have been averted), a million people starved. The term 'laissez-faire', however, can be misleading, as day-to-day economic mechanisms were actively protected and enforced by state services. During this period, between forty and seventy shiploads of food left Irish ports daily under the protection of British police, militia and army forces supported by coastguard vessels and gunships. In addition to deaths by starvation, a further million were forced into exile.

Keeping in mind that the population of Ireland in 1841 was as high as 8.5 million, one struggles to comprehend the terrible impact this cataclysm had on the long-term development of the country. As recently as 1961, the island recorded half of this — a mere 4,25 million[45] as the population figure. Many questions remain unanswerable. It is impossible to assess the loss — creative, cultural, social — that this act of genocide brought upon the country? It is in this context that the preservation of authentic Gaelic myth, culture, and iconography becomes highly significant for our current understanding of not only our past but of who we are today. Indeed, all these themes — dispossession, language, cultural loss, and the construction and the forced implementation of new identities — are all powerfully resonant in Brian Friel's classic play, *Translations*.

The considerable success of the extended English/British colonial project in Ireland accounts for the aforementioned ubiquity of neoclassical iconography, colonial statuary and royalist nomenclature we can observe today. In many respects, contemporary Ireland is a palimpsest of numerous colonialist narratives that have obscured older Gaelic-mythological narratives. This accounts for the extraordinary level of unfamiliarity, lack of awareness, and even ignorance of our ancient traditions and mythologies among our own population.

**IV.B. Colonialism as Patriarchal Dominance: Sheela-na-Gigs**

Colonialism does not only seek to control the mythologies, lands, languages, and resources of the colonised. It is also a patriarchal force that has little regard for aspects of matriarchy (either in its mythological manifestations, or simply in the traditional roles women may have enjoyed in past communities). And it must be pointed out that another patriarchal force needs to be mentioned here, as we are dealing with the role of ancient matriarchal structures, female iconography, and indeed women in society, both past and present. This patriarchal force is the Church (both Catholic and Protestant), which has acted along very similar lines to the political colonialist entities discussed above. In relation to the suppression of matriarchal elements in Ireland, one example may be brought to our attention—the case of the Sheela-na-gig. It is worthwhile invoking the Sheela-na-gig within the context of this paper, which focuses on the current collective memory of ancient Gaelic mythologies.

Sheela-na-gigs are stone-carved representations of female figures portraying notably enlarged vulvas and (occasionally) oversized heads. Contrastingly, the rest of the body is often withered, emaciated or even skeletal, exhibiting flaccid, diminutive or missing breasts. The figures are often balding with wrinkled foreheads and V-shaped grooves on their faces. Other features include exposed ribs and gritted teeth. One of the most fascinating aspects of the Sheelas is the range of attitudes they reveal—from defiant and aggressive to humorous and whimsical (especially when dancing). Their emphatic stare is nearly always direct and intense.

---

[45] In 1961 there were 2.8 million in the 26 counties of the Irish Republic and 1.45 million in the North.

Sheela-na-gigs are not mentioned by name in written records until about the mid-1800s after the Ordnance Survey of Ireland had been initiated by the British in 1825. Since then, anthropologists, archaeologists and historians have offered numerous theories pertaining to their potential meaning and function. Among these is that they are icons of fertility that facilitate conception and childbirth. Another is that they are warnings against lust or the transgression of religious taboos. Opposing these propositions is the notion that they are Celtic Goddesses representative of female empowerment and untamed sexuality. They are also seen as symbols of the power of nature to give and take life, as defensive talismans against the evil eye, or as emblems and facilitators of sovereignty over land or communities. The liminal location of many Sheelas (above church doorways or windows) and their occasionally concealed (or semi-concealed) placement within castles and church walls suggests that, for some, they had an apotropaic function (intended to ward off evil) or that they facilitated rites of passage from one state to another—from life to death, a spiritual transformation, or an elevation to higher social or political status.

Sheela-na-gigs are found on medieval castles and churches, and city and town walls; others are situated near ancient wells. About one hundred and twenty figures are recorded in Ireland and around forty have been uncovered in Britain (often on or close to monastic sites). Their placement above church doorways or close to wells (often situated on ancient roads) suggests that they were in some periods venerated. In other contexts, their intentional destruction tells a different story. It has been difficult therefore for archaeologists to establish for certain when, how, where or why they originated. While there is much disagreement among commentators, many archaeologists propose that they date from between the 11th and 17th centuries. This date frame seems to be unlikely. There is a distinctly credible hypothesis that Sheelas were placed on church doorways as a means of attracting into Roman Christianity those still engaged with residual pagan practices—the powerful female icon of a pagan communal society becomes the powerful female icon of the 'Mother Church'.

It is important to note, however, that all these plausible meanings need to be understood as a collection of individual projections and circumscriptions that have been imposed on Sheela-na-gigs, as it were, post factum. And because they are individuated perspectives (both temporally and thematically), and that they are instigated from outside the socio-cultural paradigms within which the Sheela functioned, they are likely to be, at best, incomplete ascriptions. Such singularized projections cannot account for the holistic character of the collective socio-cultural and ritual-symbolic milieus that created, nurtured and retained these 'icons'; they can't offer credible evidence of their deep-structure functionality and semiotic orientation within such archaic social contexts.

This 'naming' of what or who the Sheela is provides us with an example of the manipulation of the role of the female (be it icon or real women) by individuals and institutions that seek to control her social roles and her untamed sexuality. The Sheela-na-gig, in the ways in which she has been assigned designations by others, provides us with a perfect example of the historical treatment of women politically, socially, and even artistically (the controlled, 'beautiful' nude, nearly always painted by men). However, the abject nature of the Sheela—the way she stares back, the way she smirks, the way she exposes her vulva, the way in which she interrupts the historically controlled female form by turning herself literally inside-out to expose her dangerous and 'abject' innards, serve as an extraordinary act of political agency and feminist militancy. Historically, this act of feminist militancy has nearly always been detrimental to the female— think of Antigone, Madame Bovary, Carmen.

Julia Kristeva is valuable here in animating the role of the abject in the context of subject-object power relations, or what she calls the 'I/Other, Inside/Outside' binary [Kristeva, 1982]. This is a construction whereby subjects (men, patriarchal institutions) enjoy all aspects of agency, power and autonomy, with the further capacity to reject, repulse and oppress the 'abject other'. This agency lies in stark contrast to the abject's marginalization and powerlessness. However, it is the Sheela's very refusal to accept this constitutive power differential that makes her so significant as activist. Her agency resides in the fact that she does not represent the female body as a powerless, static entity, gazed-upon and preserved through socio-cultural and artistic power structures. Rather, she inverts establishmentarian notions of beauty; and through the blatant exposure of her orifices, she turns her body inside out to shatter the boundaries between a safely managed exterior and a corrupting, abject interior.

But she does more than just represent the abject, which would merely conserve a passive and perhaps even a submissive relation to the subject. By actively responding to the subject, by staring precociously back at his gaze, and by audaciously exteriorizing her secret and abject interiors in an act of extraordinary defiance, she weaponizes her abject body against the subject and his hegemonic constructs be they colonial, religious-institutional or socio-cultural. To recall a Kristevan image, the Sheela, by exposing her insides, gives birth to herself…she literally abjects. The vulva—in the way the Sheela turns it inside out—is skilfully recruited as an agent of political-feminist activism. And in doing this, she rejects centuries of imposed androcentric constructions of female beauty, not least contemporary versions with their rampant messages of body fascism so ubiquitous today.

To the Church, the Sheela-na-gig was merely an icon, a representation of the female that must be suppressed, as it worked against both the Augustinian strictures that underlie Catholic edict. To the colonialist, she was an emblem of ageless Gaelic traditions that must be eradicated to dissolve mythic memory, and to make way for new 'civilized' narratives that serve the occupier. But the Sheela-na-gig, as an icon represents an ancient aspect of Irish-Gaelic culture. She is an extant example of what our ancient mythological past looked like. And indubitably, she is evidence of a much stronger agential role that women played in our ancient social structures. We cannot be certain what exactly her role was, or roles were. But she is very explicit as to the role of her body. She may represent what Mikhail Bakhtin describes as the 'biocosmic cycle of cyclic changes'—wherein the body reflects the successive phases of nature ('changing seasons: sowing, conception, growth, death'); hence the Sheela's exaggerated features of sexual fecundity and birth in the swollen vulva, and of decay and death in the exposed ribs and emaciated breasts. As such, she also represents the holistic relationship Gaelic communities had with the land and the environment before the colonial project altered it irrevocably.

Why would we want to ignore the potential value of the Sheela-na-gig to our understanding of our past, indeed to our understanding of who we were, and thus who we are? This is why we need to persevere in 'unconcealing' our past mythologies, to bring back into the open an understanding of the role of women in Gaelic-Irish societies of the past. This is why most of our rivers were named after female Goddesses. This is not a trivial matter. The long colonial project discussed here has impacted deeply on our collective consciousness. It has attempted to airbrush out the deepest aspects of who we are — our mythology, language, music, Gaelic social structures, Gaelic legal apparatuses, etc. But colonialism — be it in the form of political or religious-institutional dominance — has also had a detrimental effect of our understanding of the role of women in our mythic memories, and thus in our contemporary Ireland.

# V. ÉIRÍ: Évoking Ireland's Resilient female Ícons

Having established how deep the colonial mindset is in Ireland today, its origins, and a public desire to resist it, and inspired by the success of "Arts for Sinann", we next use similar methods to those used in Section II to invoke more of Ireland's resilient female icons. To this end we ran a second competition – but this time not centred on a single on a single goddess but on any – including Sheela-na-gigs!

We invented the acronym ÉIRÍ to stand for Évoking Ireland's Resilient female Ícons. It also played on an Irish word meaning "stands up", arises", "ascends", "soars." ÉIRÍ was an international arts competition in parallel with a participatory research project aimed to re-ignite awareness of inspirational female figures from Irish mythology and folklore "who have all too often been forgotten, suppressed, or overwritten in the mists of time." Again, members of the public were invited to submit original art in any form, but this time inspired by any female figure. People were also invited to send their own research into women in local or national mythology or folklore whom they feel have been forgotten or underrepresented.

The hope was that this project may lead to better representation of women in Ireland's national iconography and identity documentation. Despite the country taking its name from the female mythological figure Éiru, and despite the majority of Ireland's population being female, all but one of the 14 figures in modern-day Irish passports are male. This is because these too are taken from the Custom House in Dublin – they are the same neo-classical Ossianic, constructs that we spoke about earlier – the same iconography as used in Athlone, iconography that represents commerce instead of culture, profit instead of people, and men instead of parity.

ÉIRÍ was on a bigger scale with €10,000 in prizes (including up to €3,000 for schools). In recognition of the research achievements of MMM, and to support impact of its research on the world beyond academia, Coventry University provided about half the funding for this competition. The other half came from the University's allocation of funding for Participatory Research from Research England. Apart from the larger scale, ÉIRÍ ran similar to "Arts for Sinann". The "Calibrate with Confidence" algorithm was again used to account for differences in stringencies and confidence levels of assessors. The big difference between ÉIRÍ and Sinann was that for the new project we invited the Celtic Eye Art Group. This is an artistic outfit working with the Karst Farming Group of South Roscommon to highlight the importance of biodiversity in protecting the landscape of the area. Celtic Eye's mission is to highlight folklore, myths, stories of our land, and biodiversity and to bring people together in a positive way through art, poetry, music, dance and other art forms. Celtic Eye wis very successful at lobbying and has recently been invited to ed to the Seanad Éireann in recognition of their work. Therefore, they represented a powerful fifth element.

The project was announced in *The Irish Post* on 14 May 2022[46] with deadline for submissions of 31 October. This is the date of Ireland's major pre-Christian festival which gave birth to Halloween (and has its origins at Oweynagat in Rathcroghan). It marks the end of the harvest season and, since we aimed to harvest insight into Irish mythology and artistic representations of it, we deemed it an appropriate deadline.

**V.A. Informing the People**

Again, we had a series of feature articles in *The Irish Post*. The first was by Mal Rogers, one of the judges in the Sinann project and now Editor of the newspaper[47]. The feature gave details of the competition and the philosophy behind the project. This was followed on 28 May[48] by a quote from highly respected Celtic scholar Peter Beresford Ellis who viewed Irish mythology as one of the brightest gems of European cultural inheritance. "It is both unique and dynamic, a mythology which ought to be as well-known and as valued as those of ancient Greece and Rome." "So why is it not?" the article went on to ask, "And why are Ireland's passports full of neoclassical iconography? Why are all but one of its "river gods" men when half of the passport holders are women?". The tempting article invited readers to "Follow the pages of this newspaper over the coming months to delve into Irish mythology and women's roles in it."

---

[46] "Help reawaken Ireland's past, *Irish Post*, 14 may 2022
[47] Mal Rogers, "Competition news – with women in the leading role", *Irish Post*, 21 May 2022
[48] "Competition news: 10,000 to be won", *Irish Post*, 28 May 2022

In early June came the next item in *The Irish Post*[49] which repeated the questions posed a week earlier. The item went on to allude to an account in the newspaper in 2016 (repeated online on 1 July[50]) which reported Yose's et al's original paper [Yose et al, 2016]. In the 1760s, the item said, James Macpherson claimed to have translated the epic poems of Ossian from Scottish Gaelic. "He distanced them from Irish sources, claiming they were like the Greek Classics. Controversy arose as Irish scholars immediately saw them as misappropriating Irish mythology…. They found a remarkable similarity between Ossian and Irish mythology – and little similarity to the classics." (See Appendix A.)

On 11 June Chris Thomson wrote her second piece for *The Irish Post*.[51] "You may have heard of the precocious boy hero, Cú Chulainn, the canny Fionn Mac Cumhaill, or even memories of Manannán of the waves", Thompson wrote, but "what about women characters?" Thompson went on to give a brief outline of some of the aspects we associate with characters such as Medh, Niamh, Scáthach, Fuimnach, Etain, Dubh Lacha, Emer, and more.

> These women are no supporting cyphers, also-rans or 'proto' Disney princesses. These are real women. They are inspiring, thought provoking, adventurous, occasionally disturbing, and often funny. They make choices, both good and bad. Medb is a good example. In her bid for the leadership of Connacht, she is said to have drowned her sister, an action that eventually, led to her own death. Yes, they make mistakes and live with their choices.
> These are women who lead. They advise, they record and report events. They matter. Their stories will still resonate with women today and deserve to be far better known.

A week later[52] Benjamin Dwyer told the readers of *The Irish Post* about Sheela-na-gigs: "hidden away for centuries they are now resurging with ÉIRÍ", he said. On 2 July, Joe Horgan did a column about Irish diaspora in England and elsewhere.[53] His full-page item focused on how the Banshee, "a harbinger of death" followed the emigrants. A week later came another item from Rathcroghan – this time with Daniel Curley joined by Mike McCarthy.[54]

Following a brief description of the archaeological significance of the Rathcroghan landscape, McCarthy and Curley described some of the mythology associated with it. In particular, the "Warrior Queen Medb" was brought to the fore, as was the Irish Battle Goddess, the Mórrígán as well as Ériu of the Tuatha Dé Danann. "When this goddess met the Milesians at the Hill of Uisneach, the poet Amairgin promised her that the country would bear her name, and this legacy has passed down through the generations." The similarity between ÉIRÍ and Ériu is no coincidence for, by using the former as acronym for the project we hoped to tap into the national significance of the latter (it is from this that Ireland takes its name in Irish – Éire).

---

[49] "Awakening Ireland's ancient mythology", *Irish Post*, 4 June 2022;
https://www.irishpost.com/news/delving-into-the-female-side-of-irelands-history-235538
[50] "The famous Scottish tales that are actually Irish", *The Irish Post*, 23 July 2022;.
https://www.irishpost.com/culture/the-famous-scottish-tales-that-are-actually-irish-236610
This item was repeated in print and was first printed in https://oldmooresalmanac.com/famous-scottish-tales-actually-irish.
[51] Chris Thompson, "Evoking Ireland's resilient and iconic females: ÉIRÍ", *The Irish Post*, 11 June 2022.
[52] Benjamin Dwyer, "The female side of mythology", *The Irish Post*, 18 June 2022.
[53] Joe Horgan, "Legendary women, real and mythical", *The Irish Post*, 2 July 2022.
[54] Mike McCarthy and Daniel Curley, "Connacht's archaeological and mythological treasures", *The Irish Post*, 9 July 2022.

On 16 July *The Irish Post* ran a consolidating feature "Maths Meets Myths" by Kenna.[55] The purpose of this article was to reaffirm the mathematical origin of the ÉIRÍ project that, by now, so many *Irish Post* readers and *Story Archaeology* users were engaged with. To make the connection, Kenna invoked the nephew of a swash-buckling officer in the Irish Brigade of the French army in the 18th century. "Duelling Dick" has also served in the army of the empress of Austria and had the right appeal to catch the eye. His nephew Richard Kirwan served Ireland in different ways - both as a United Irishman and as a physicist. The United Irish movement aimed to overthrow the protestant ascendency to unite Catholic, Protestant and dissenter into a new egalitarian Irish society. As a scientist Kirwan's aim was to advance the welfare and prosperity of Ireland's population.

As noted in Kenna's *Irish Post* piece, one of Kirwan's many contributions was to teach us that if there were no interactions between atoms there'd be no magnetism. Likewise, a society is formed by interactions. And here comes the connection to mythology; the society that is described in a narrative has to have interactions to keep it intact. Large amounts of atoms interacting to make a magnet is not unlike large amounts of people in a society or large amounts of characters interacting in stories to make a mythology. This is the connection that had been made in 2010 that formed the MMM project.

After describing the connection between maths and myths, Kenna went on to reinforce the aim of the ÉIRÍ project and questioned again the usage if 18th century (male) iconography in modern-day passports. He pointed out that the stories ÉIRÍ invokes involve both women and men and predate Catholic, Protestant, dissenter "and everything else" and would form an ideal basis for identity documents.

The *Westmeath Independent*[56] similarly vented Kenna's concerns about the Irish passports.
> That is one of the reasons why, when Ireland's passports feature 14 neoclassical images on the inside pages, all but one of them are male', he says. He contends that had Irish mythology been used instead of neo-classical iconography, Ireland would have more gender-balanced passports.

And so it rested in the minds of the people over the Summer months. The next burst of media activity came in Autumn with an article by renowned Irish artist Nicola Bowes.[57] Bowes explained how iconic figures like Queen Medb led her "to become involved with a biodiversity, heritage, and folklore project in collaboration with the Karst Farming Group of South Roscommon." As she explained, "Karst is a topography characterized by underground drainage systems and caves" and "Such fragile ecosystems need to be preserved". Bowes "knew an ambitious project would need diversity of experience to highlight its importance" so she founded *Celtic Eye Art Group* to include artists, poets, writers, herbalists, dancers, and farmers. The Group rapidly grew to 1,400 members and holds exhibitions and display diverse art from all over Ireland.

---

[55] Ralph Kenna. "Maths Meets Myths", *The Irish Post*, 16 July 2022.
[56] Geraldine Grennan, "Athlone academic launches international art competition", *Westmeath Independent*, 3 September 2022
[57] Nicola Bowes, "Inspired by Irish Mythology", *Irish Post*, 24 September 2022.

Bowes went on to explain the significance of characters such as "Clíodhna, Queen of the Banshees of the *Tuatha Dé Danann,* the supernatural race who, while dwell in the "other world"." She also wrote about Brigid who from 2023 would be recognised in Ireland with her own bank holiday. Bowes item goes on to interlace the healing powers of mythological and historical women. As with the former the latter too were "played down in male-oriented historical literature."

> My research into these females was as invigorating as the art and poetry they inspire. This is why, when I was invited to be part of the very important *ÉIRÍ* project, it felt like our goddesses are indeed rising. For now it is time that iconic females are recognised for the very important roles they play in protecting our word - its past, present and future.

The media campaign continued in September with another item from Rathcroghan.[58] The focus was again on Queen Medb because of her deep associations with that Royal site. As reported, "the literary character of Medb looms large in the Irish psyche. She is the great leader of the Connacht armies in our national epic tale, An Táin Bó Cúailnge, the Cattle Raid of Cooley." The question asked was whether she is "a flesh and blood warrior queen, or is she perhaps a female divinity in human form?"

After a tour de force about other goddess figures in early Europe and Ireland, the authors described the many equine associations of Medb. They describe how, in early Ireland, "it was these female divinities that the king had to symbolically 'wed' and this 'marriage' known as *Banais Ríghi* or a 'Kingship Marriage' served as a link between his realm and the divine."

> By this divine marriage, the king acted as a bridge to the Otherworld, and so the health of the kingdom, particularly the harvest, was dependent on the king's own good condition and behaviours. Should the union be unrighteous, particularly in early historic times, there is suggestion that the king could be replaced by a more suitable candidate, and the failed king was sometimes disposed of in a most brutal fashion.

As pointed out by McCarthy and Curley, when we consider Medb in this light, some tantalising aspects emerge; while the kings were transitory, Medb was constant. So perhaps the literary character of Medb is in fact an echo of "a very important and significant prehistoric, primordial goddess of territory in Ireland

---

[58] Mike McCarthy and Daniel Curley, "Divine Women of Ireland", *Irish Post*, 24 September 2022.

who bestowed sovereignty on the king through the *Banais Ríghi* — Sacred Marriage ritual."

This series of articles ended on 8 October when Andy Kilmartin, a London-based Irish educationalist, described how the spirit of the Irish goddess Ériu lives on in Chile and encouraged Ireland's global diaspora to partake in the ÉIRÍ project. And on 29 October, with the ongoing war in Ukraine attention turned to that country and how their national epic compares to Ireland's Táin Bó Cúailnge.[59] In parallel with all of the above coverage in printed press, *Story Archaeology* had carried out a number of broadcasts which are still available. Kenna was able to further promote on Athlone community Radio[60].

**V.B. Évoking Ireland's Resilient Female Ícons (ÉIRÍ)**

And thus ended the vast amount of printed-media coverage of ÉIRÍ, carefully timed to educate the public over the course of the competition. Following short items on 11 February[61] and 20 May[62], the results of the adult category were announced in a two-page spread on 27 May 2023.[63] Those for the schools category were release on 24 June 2023.[64] We display those from the Adult category in Figure 6.1.

| 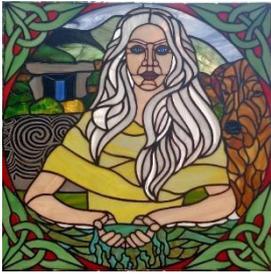  **"Boand/Boann"**  After reading some old Irish poetry mentioning Boand that was translated into English, I garnered an image of her being fair skinned from phrases such as "white bright Boand", "fair formed" and "white breasted Boand". This resulted in my use of white glass for her hair and a light skin tone glass.  Medium: Stained glass (350 pieces, 20 inch x 20 inch)  Lee Fenlon | 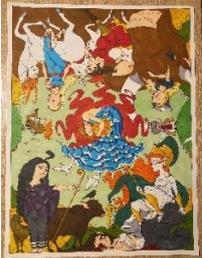  **A multitude of Irish female mythological and folkloric figures**  Medb, Macha, Naoimh, Caileach Beara, Morrigan, Oengus and Caer Imbormith, Deirdre, Eriu, Tailtiu, Boann  Medium: Drawing  Daniel Breheny | 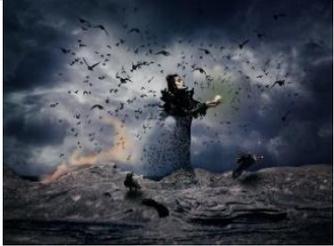  **"Morrigan"**  To capture the moment of change from her human to her crow form, a symbol that I see with her. I wanted to show her strength and assurance in her role as a recorder of the past and the force that brings in the new. It was important that she was not a malevolent or a kind, benign being. She is an active force transcending good or evil, calmly recording and letting you know what will happen if you continue of this path. For me it is a personification of karma.  Medium: Photograph & digital  Kate Lionis |
|---|---|---|

---

[59] "Heroes and Legends", *The IIrish Post*, 29 October 2022.
[60] Athlone Community Radio, interview with Ralph on 01/09/2022 (broadcaster = Deirdre Ó Murchadha).
[61] "Competition update", *The Irish Post*, 11 February 2023
[62] "The ÉIRÏ arts competition", *The Irish Post*,, 20 May 2023
[63] "Eyes on the ÉIRÍ prize", *The Irish Post*, 27 May 2023
[64] "Schools embrace mathematician's art project", *The Irish Post*, 24 June 2023.

| | | |
|---|---|---|
| 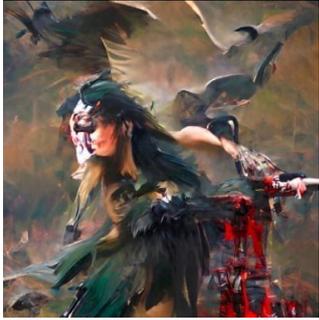 | 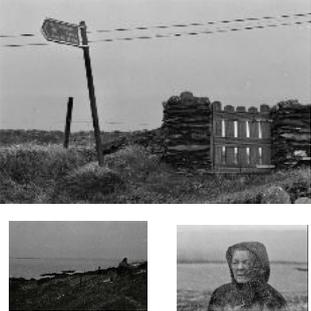 | 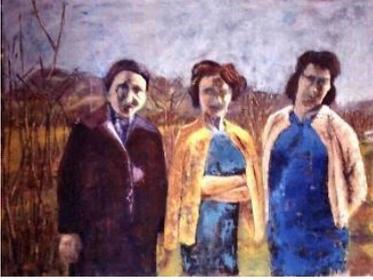 |
| **"Morrígan"** | **"Chailleach"** | **"Morrigan"** **"(Macha, Badbh & Nemain)"** |
| This is the A.I.'s understanding and interpretation of the Morrígan, the goddess of war, fate, and destiny. It's full of movement and really encompasses the Morrígans shape-shifting abilities. She looks as though she is in the middle of a pitched battle, looking off to the side of the image with a ferocious intention. The AI has no agency, no will, it is simply used to elevate human creativity. Medium: Digital | I used the complex process of double exposure in which two images are exposed to the same frame of film, chemically binding them. I feel this was the best way to capture and demonstrate the bond between a place and its people, a landscape and its myth. Medium: A number of photos with accompanying text. | The Morrígan in Irish mythology is often depicted as a shape-shifting goddess: one of three sisters Macha, Badbh and Nemain; or a goddess of the Tuatha de Danann with three elements to her. This image can be seen as a portrayal of the spirit of the three women, or elements, who form the Morrígan as she shapeshifts down the generations through the normal day-to-day lives of Irish women. Medium: Painting |
| Louise Shine | Jesse Downs | Ruth Egan |

**Sheela-Na-Gig Triptych. Three poems: "The three main theories of Sheela-na-gig"**

| Goddess | Apotropaia | Sermon |
|---|---|---|
| Fecundity - lush, fertile word for the arable land found in a womans belly. Ready to be ploughed by plunging ploughshare, the rich red soil turned, ripe for seeding. Blank eyes, fingers pulling herself apart, offering a glimpse into her fossil body. Crude, in design and by design, an ancient face that saw the rivers flood a thousand times, a periodic shedding of what was there before. | Don't think you can get by me… I am more powerful than you can know. Look at what I have between my thighs. Look! Look away… there is magick here. I can calm the sea, tame whirlwinds, catch lightning. That which brings life will bring you death… the Romans knew this. Beware - devils, frail deities, look on this with shock and awe… don't think you can get by me… | Sin made solid - lust delineated. See how she shows herself, see how her sex insinuates your soul, weak flesh turned stone. Gorgon-glanced and scrape-sculpted, hag breasts and rictus smile, see how your God punishes. Blank eyes, fingers pulling herself apart, scalpel scarred, scarified. See her succumb to Satans whisper, watch her sink into his seventh circle. Nailed above Gods door for a thousand years, silently staring still. |

Alexander Goodison.          Medium: Poetry

**Figure 6.1:** Winning submissions to the ÉIRÍ competition.

There were over 150 submissions to the ÉIRÍ competition. The 80 shortlisted artists had to prove they did their own research into the characters they depicted. This is an extraordinary breadth of engagement and the quality of submissions (winning entries are depicted in Figure 6.1) demonstrated depth. The panel comprised 28 people and, between them, they supplied 410 scores altogether (5.125 scores per item). This is much more manageable than what we would have had if everyone had to assess everything, in which case we'd have had 28x80 = 2,240 scores. The CWC algorithm aims to achieve the results we would have got had everyone assessed everything by properly calibrating scores for different degrees of stringency and confidence.

The panel comprised folklorists, mythologists, archaeologists, political scientists, professional artists, musicians, musicologists, sculptors, film makers, teachers, application developer, engineers, professional writers and journalists, managers, mathematicians, physicists, geographers, date scientists, complexity scientists, statisticians. These were 11 scientists, 10 humanities people, 7 lay people: 15 males and 13 females. Our panel included professors from universities in Limerick, Belfast, Berkeley, Warwick, Princeton, Middlesex, London, Coventry as well as the Institute for Condensed Matter Physics, Lviv in Ukraine and University of the Sunshine Coast in Australia.

Again unsolicited comments from the artists ensued. These included:
- First of all, I would like to say what an interesting project this has been, I have learnt so much while researching for it and have enjoyed it immensely. I thoroughly enjoyed making and researching this project and I hope you enjoy my contribution.
- I believe the work you are doing is incredibly important to the cultural heritage if Ireland and I thank you for it.
- The project is such a brilliant idea and I am looking forward to reading and seeing the results from all the contributions you have encouraged in the coming months
- It was an absolute pleasure to research and make this piece for the competition. I'm fascinated by the stories of women in Irish Mythology.
- It was enjoyable trying to design and create this work for the competition, and got me to think about the mythology in a new light.
- My success in the An Sinann adult art competition opened up a new world for me and my artistic career.

This time we asked the judges for a few comments too:

- "…my commentary is leading me down some fascinating roads and is more time-consuming than originally envisaged. This is not in any way a problem: in fact, I'm quite fascinated by the subject, having never considered the range and scope of Irish mythology."
- "One of the most interesting things I have done for a long time! Many thanks for the invitation."
- "It's been really interesting to look at/listen to/read everything. It has been quite difficult to judge though. I hope I've been fair, and thank you for including me in the process."
- There were really some impressive contributions among the submissions!! Really enjoyed that! Would love to see all of them.
- I would be a strong believer that anyone who undertakes any creative endeavour and then submits it for public perusal deserves the highest admiration. It is easy to sit on the fence, but putting yourself out there always takes bravery and therefore I tried to be as positive as I could and look for merit in every entry. I hope I have reflected this in my accompanying commentary. I felt the standard overall was very high, and three images in particular, although all completely different, stood out in my opinion. Thanks for the experience, I really enjoyed it and if you require clarification on anything, please let me know.
- It was interesting to learn about Irish mythology and even more interesting to learn how different people not only concentrated on different parts of it but also interpreted it differently. And I must admit your judging system made sure I paid a lot more attention to both the notes from the artists and what you were looking for.

- Thanks for including me in this! I spent the afternoon going through these and trying to give them a clear score. I found it very hard to judge the stories; you'd think as a Humanities professor I'd have had an easy time of those...
- Thanks for involving me, I enjoyed looking at the submissions, some great stuff in there.
- I am impressed and inspired by the talent, creativity, effort, and wonderful artwork submitted. It was a great pleasure looking at this.
- And can I just say that I'm finding this quite fascinating. Both the art work and the judging process.
- They are quite incredible. Really convey the three theories in a very striking manner - so much so we'll never forget them! Our own interest in Sheela-na-gig's has definitely been inspired by this work. Remarkable how three contradictory interpretations can be expressed so vividly and persuasively.
- Thank you very much for involving me. It was really quite fascinating, and the work was really impressive.
- I learned that I think I am a more generous marker than everybody else. I will be interested to see if that's true!

Thus, we come to the end of the ÉIRÍ participatory and arts project. Its aim was to evidence that centuries of colonialism was not 100% effective and (a) people remember and even though not formally educated in the topic instinctively know the value of Irish mythology and women's roles in it and (b) Ireland has sufficient talent to depict her heritage in any artistic medium.

# VI    Conclusions

In this paper we have recounted the extraordinary story of the rise of Sinann – the figure who was recently overwritten by authorities and forgotten by the people. We explained how the Maths Meets Myths project exposed an incoming statue that claimed to be the "river god head" of the Shannon as a colonial male imposter. Once informed, the local people protested in the form of letters to newspapers, street events and a petition. Although the neo-classical, colonial statue was erected it is now despised by townsfolk while Sinann is loved.

We then went on to report an arts competition designed to instil awareness of Sinann as sovereign over the river that bears her name. After that we recounted how remarkable instinctual insight led to delving into the source rather than relying on renditions that were tailored for Victorian times. An authoritative fresh translation led to an altogether different Sinann – "no supporting cyphers, also-rans or 'proto' Disney princesses" but a powerful figure in her own right.

We then examined the origins of colonialism and its presence in modern-day Ireland. We saw how the erasure of Gaelic cultural and mythology was important for the colonizer who saw them as repositories of Ireland's past. We also saw how colonialism brought patriarchal forces which have scant regard for matriarchal values that were prominent in Ireland in ancient times. We used Sheela-na-gigs as an exemplar of defiant female icons of Irish-Gaelic culture who have weathered the storm – an example of what our ancient mythological past might have looked like.

We then escalated to promote awareness of any or all of Ireland's iconic females. This was the ÉIRÍ arts project which was run on an altogether bigger scale to the local Sinann one. Through both art projects we have demonstrated that Ireland has sufficient talent to produce well-researched, high-quality products that can draw from Irish-Gaelic culture.

We hope we have established that mythological stories were not intended merely as light entertainment. Indeed, they encoded cultural information including warnings about the consequences of breaking 'natural' law. This is certainly true of many Irish stories. The underlying thread of the *Táin Bó Cúailnge* for example warns that if natural law, particularly concerning women and childbirth, is set aside then prosperity would be lost, and the land laid to waste. These stories were seen as a threat to the colonial mission that devastated Gaelic mythology and identity.

Patriarchy and colonialism went hand in hand down the centuries.[65] And they left a legacy. Fast forward to the 20th century and we have persecution of women by church and state in the form of so-called Mother and Baby homes. These were institutions in the 1920's up to as late as the 1990's, where unwed women were sent to deliver their babies. Investigations in the 2010's found that around one in seven (9,000 children) children born there between 1922 and 1998 died – double the infancy mortality rate in the general population. An investigatory report was published in January 2021 when the Taoiseach (Irish Prime Minister) made a formal apology to survivors on behalf of the state. This was far from satisfactory however and criticise by survivors as "deeply flawed".[66]

Now, as we face different challenges, our old stories can, again, provide a reminder that if we abuse our people and our environment, it will become a wasteland. It is time for the original and oldest stories, with themes of hope in a time of change to become spoken once more. And indeed, they are. The Celtic Eye Art Group is a prime example. Connecting farmers and artists their aim is to highlight and protect the unique landscape[67] of South Roscommon in Ireland. Their community of supporters is vast, reaching artists, poets, musicians, herbalists and many more and sporting significant online presence. They use Irish mythology to make real the powerful connection of Celtic goddesses with the land.

We conclude as we started, with Sinann — by reproducing some more text from Thompson's 2020 interview with *The Irish Post*.[68]

---

[65] In comparison to other European countries, for example "Witch trials were rare in Ireland, a culture where ancient beliefs in powerful female archetypes remained." It was Scottish setters in the 1600's who brought their prejudices with them. [Paula Kehoe, in "An Aíabhal Inti (The Devil's in Her)", *BBC Gaeilge* and *TG4*, broadcast, 16 October 2022.]
[66] "Too Little Too Late", *Irish Post*, 23 January 2021.
[67] Karst is a topography characterized by underground drainage systems and caves, another Irish example being the Burren in County Clare. Such fragile ecosystems need to be preserved and that is the aim of the Group.
[68] Chris Thompson, "Celebrating Sinann", *Irish Post*, 26 December 2020; https://www.irishpost.com/life-style/why-the-story-of-the-goddess-of-the-river-shannon-is-one-worth-telling-200145

We live in a time of change, of uncertainty. With a pandemic, economic struggle, climate change and war all around us, we might be living in a dystopic Sci-Fi film, and yet, this is all happening for real. This is a time that cries out for inspiration, imagination, enthusiasm, new ideas and, above all, hope.

It is a moment in history requiring people of vision - artist, scientists, health workers, teachers, everyone - to reach out for a source of new possibilities and discover unexpected ways of working.

The story of Sinann is able to remind us that there is always an inspirational source in that Otherworld of imagination and creativity.

Tough challenges may well lie ahead but, so the story tells us, there is a well with the nine hazel trees of wisdom that flower and fruit at every season of the year. It is a well that cannot dry up. It is the well of imagination, vision, and creativity. The story of Sinann still offers us something to celebrate and the River Shannon will always remember her.

"Olen dia sruth a sáer-ainm" - her noble name follows her river.

\* \* \* \* \* \* \*

## Acknowledgements

The authors wish to thank Coventry University and Research England Participatory Research for funding this project. We thank the Fiona Audley and Mal Rogers of *The Irish Post*, Tadhg Carey and Adrian Cusack of the *Westmeath Independent*, Eunan Keys of *Athlone Community Radio*, Ursula Ledwith of *RosFM* 94.6 and other printed media and broadcasters for helping communicate MMM to the people. The Sinann campaign was led by Orla Donnelly and Fiona Lynam as well as Caroline Coyle. It was these women who obtained information through the Freedom of Information act; set up Facebook forums; organised the petition; and organised street events. We thank Caroline Mannion for two letters published in the *Westmeath Independent* in support of Sinann. We thank Bob Fox, John Madden and Dom Parker for the photos used in Figure 2.4 and Paul Mulvey for that in Figure 2.5. We thank Ruairí Ó Leochain for his encouraging interview in *The Irish Post*. RK is grateful to Ian Kenneally for inviting and organising his lecture at the Old Athlone Society. We thank the artists and followers of Celtic Eye Art Group and the farmers and community of volunteers of South Roscommon in Ireland who helped promote the ÉIRÍ project. We thank the 47 artists who submitted to the Sinann competition and over 150 who submitted to ÉIRÍ. We thank the 30 judges who, beside the authors of this paper, included David Arundel; Martine Barons; Michael Butler; Colm Connaughton;

Tim and Sarah Ellis; Alison Haigh; Annie Heeney; Lelia Henry; Joe Horgan; Susanne Horn; Madeleine Janickyj; Shona Shirley Macdonald; Paul Manning; Marisa McGlinchey; Olesya Mryglod; Patrick Nunn; Arij Al Soltan; Timothy Tangherlini; Noel Tighe; and Nataliya Yanchevskaya. We thank the 127 people who signed the open letter and the 700 who signed the petition. Finally, we thank Sinann and the inspirational female icons and all those who throughout the centuries have done enough to preserve their legacy.

# References

Bartlett T. 2010. *Ireland: A History*. Cambridge: Cambridge University Press.

Cronin A. 2002. "Ireland Under the Union: A Backward Look," in *Hearts and Minds: Irish Culture and Society Under the Act of Union*, ed. Bruce Stewart. Monaco: Princess Grace Irish Library.

Dooley A. and Roe, H. 1999, *Tales of the Elders of Ireland: A New Translation of* Acallam na Senórach. Oxford: Oxford University Press.

Fanon F. 1968. *The Wretched of the Earth*. New York: Grove.

Gaskill H. (Ed.) 1996. *The Poems of Ossian and Related Works*. Edinburgh: Edinburgh University Press.

Gregory A. 1904. *Gods and Fighting Men: The Story of the Tuatha De Danaan and of the Fianna of Ireland*, trans. Lady Augusta Gregory. London: J. Murray, 1904; reproduced in Lady Gregory's *Complete Irish Mythology* (Vacaville, CA: Bounty Books, 2004).

Gwynn E. 1906. *The Metrical Dindshenchas*, Second reprint [x + 562 pp. words]. Dublin Institute for Advanced Studies: Dublin (1991) (reprinted 1941); https://celt.ucc.ie/published/G106500C/index.html

Hastings A. 1997. *The Construction of Nationhood*. Cambridge: Cambridge University Press.

Kane B. 2011. 'Languages of legitimacy? *An Ghaeilge*, the earl of Thomond and British politics in the Renaissance Pale, 1600-24', in Dublin and the Pale in the Renaissance. Dublin: Four Courts Press.

Kristeva J. 1982. *Powers of Horror: An Essay on Abjection*. New York: Columbia University Press.

Mac Carron P. and Kenna R. 2012. *Universal properties of mythological networks*, Europhysics Letters Vol 99 pp 28002. DOI 10.1209/0295-5075/99/28002

R.S. MacKay R.S., Parker S., Low R. and Kenna R. 2017. *Calibration with confidence: a principled method for panel assessment*, Royal Society Open Science volume 4 pp 160760; DOI: 10.1098/rsos.160760.

Macpherson J. 1760. *Fragments of Ancient Poetry, Collected in the Highlands of Scotland, and Translated from the Galic or Erse Language* (Edinburgh: G. Hamilton and J. Balfour, 1760); *Fingal, an Ancient Epic Poem, in Six Books; Together with Several Other Poems, Composed by Ossian, the Son of Fingal. Translated from the Galic Language* (London: T. Beckett and P.A. DeHondt, 1762); *Temora, an Ancient Epic Poem, in Eight Books; Together with Several Other Poems, Composed by Ossian, the Son of Fingal. Translated from the Galic Language* (London: T. Beckett and P.A. DeHondt, 1763); *The Works of Ossian, the Son of Fingal, in Two Volumes. Translated from the Galic Language by James Macpherson. The Third Edition, To Which Is Subjoined a Critical Dissertation on the Poems of Ossian* (London: T. Becket and P.A. De Hondt, 1765); and *The Poems of Ossian, Translated by James Macpherson, Esq., in Two Volumes* (London: W. Strahan and T. Becket, 1773).

O'Connor C. 1766. *Dissertations on the History of Ireland; To Which is Subjoined, a Dissertation on the Irish Colonies Established in Britain with Some Remarks.* Dublin: G. Faulkner.

O'Curry E. and Sullivan W.K. [ed.]. 1873. *On the manners and customs of the ancient Irish: a series of lectures*, London, pp 142 – 144.

O'Halloran S. 1763. *The Poems of Ossine, The Son of Fionne MacComhal, Reclaimed"* Dublin Mag. 21–23; reprinted in Vol. III of D. Moore, *Ossian and Ossianism* (London: Taylor and Francis, 2004), 87–89.

Pine R., 2014. *The Disappointed Bridge: Ireland and the Post-Colonial World*. Newcastle upon Tyne: Cambridge Scholars Publishing.

Rieu E.V. 2003. *Homer, The Iliad,* trans. London: Penguin Classics.

Said E.W. 1993. *Culture and Imperialism*. London. Chatto & Windus 266.

Shewring W. (Ed.) 1980. *Homer, The Odyssey.* Oxford Oxford. Oxford University Press.

Spenser E. 1997. *A view of the present state of Ireland*. Oxford: Oxford University Press.

F. Stafford, "Introduction: The Ossianic poems of James Macpherson," in *The Poems of Ossian and Related Works*, H. Gaskill, (ed.) (Edinburgh: Edinburgh University Press, 1996).

Yose J., Kenna R., MacCarron P., Thierry P. and Tonra J. 2016. *A Networks-Science Investigation into the Epic Poems of Ossian*, Advances in Complex Systems Vol.19 pp 1650008.

**Appendix A: What's maths got to do with it? Network science foundation for claims of misappropriation**

Network science is the bedrock of the story we tell in the main part of this test. Yet it is not necessary for the story itself (this appendix can therefore be skipped by a hurried reader). For the sake of completeness, we give a very short resume of the essentials here. We also briefly recount the story of Ossian because it resonates strongly with what appears in the main text. Indeed, we often refer to the iconography of Dublin's Custom House, that of the Mask of the Shannon and that appearing in modern-day Irish passports as *Ossianic*.

In 2012, some of this team embraced an increase in interdisciplinary methods to apply the new science of complex networks to questions in comparative mythology. Investigations of social network structures (how society is interlinked through characters interacting with each other) embedded in epic narratives allowed universal properties to be identified and ancient texts to be compared to each other [MacCarron & Kenna, 2012]. We named the concept and the project *Maths Meets Myths*, or MMM for short, the tongue-twister reflecting seemingly odd pairing of disciplines. Over the past number of years, MMM evolved to address increasingly nuanced questions and in 2017, we applied network theory to investigate the poems of Ossian [Yose et al 2016].

In 1760, one James Macpherson published the first of a series of epic poems which he claimed to have translated into English from ancient Scottish-Gaelic [MacPherson, 1760]. He claimed the poems came from a third-century bard named Ossian and they quickly achieved international acclaim, being described as among "the most important and influential works ever to have emerged" from Britain or Ireland, [Gaskill, 1996]. Comparisons were drawn to major works of the epic tradition such as Homer's *Iliad* and *Odyssey*. However, doubts about their authenticity provoked one of the greatest literary controversies of all time. Uncertainties centred on suspicions that the poems misappropriated material from Irish mythological sources.

The comparisons made between *Ossian*, the Classics and Irish mythology prompted an invitation from our recently developed networks-science point of view. We looked at the social-network structures that are embedded in *Ossian* with those of Homer's epics and Irish mythological texts: specifically, *Acallam na Senórach* (*Colloquy of the Ancients*), from the so-called Finn Cycle in Irish mythology. Note from Section II that, although she does not explicitly appear in *Acallam,* this is the cycle associated with Sinann. Our aim in [Yose et al, 2016] was to determine whether the network structures that underlie the society depicted in Macpherson's work are similar to those of either corpus. We found that the Ossianic network is dissimilar to those of Homer but had a strong structural similarity to the society underlying *Acallam.* This suggested a clear affinity between Macpherson's works and the Finn Cycle.

Briefly, the context behind Ossian is the following. In the eighteenth century, the Highlands of Scotland were culturally different from the rest of Britain and continental Europe with its own societal structure and customs. However, in 1746, defeat at the Battle of Culloden led to its assimilation into Great Britain and the suppression of native Gaelic culture and Jacobite clans. Against this backdrop, Macpherson's first volume, *Fragments of Ancient Poetry, Collected in the Highlands of Scotland, and Translated from the Galic or Erse Language*, was published in 1760. By connecting its readers to a lost heroic age, the work was motivated by a desire to mend some of the damage inflicted on the Scottish Highlands after the Jacobite Rising [Stafford, 1996].

The fragments were supposed to have originated from Ossian, whose rhythmic tales of strife, war, and love evoked an ethereal atmosphere which greatly appealed to the public. Two further Ossianic volumes, *Fingal* and *Temora*, followed and in 1765, the corpus was collected in *The Works of Ossian, the Son of Fingal*. This included "A Critical Dissertation on the Poems of Ossian, the Son of Fingal," by Hugh Blair. We based our network analysis on the text of the recent edition by Howard Gaskill [Gaskill, 1996]. The claims for an ancient and noble Scottish literary heritage were welcomed by Scottish scholars such as Blair, Adam Ferguson, and David Hume, as a catalyst for an increased sense of national identity when confronted by the cultural fragmentation of the Highland Clearances. Scholarly works such as Blair's *Critical Dissertation*, attempted to bolster the credibility and authenticity of the Ossianic poems. Thus, *Ossian* was of crucial importance in promoting Highland traditions and culture beyond the north of Scotland.

Reaction to *Ossian* in Ireland was furious, however. It was obvious to Irish antiquarians that the poems were drawn from tales of the Finn Cycle. Thinly disguised characters, places, and situations from the Irish epic tradition were readily identified. In 1763, Ireland's Sylvester O'Halloran denounced Macpherson's deceptive attempt to misappropriate Irish culture and its Gaelic heroes for Scotland [O'Halloran, 1763]. The antiquarian and Gaelic scholar Charles O'Conor further championed the Irish cause [C. O'Connor, 1766]. For example, the character Ossian—Macpherson's "illiterate Bard of an illiterate Age" — was identified as Oisín, a warrior-poet of the Finn Cycle in Irish mythology. Ossian's father, Fingal, who was a third-century Scottish king in Macpherson's texts, was identified as Fionn mac Cumhaill, leader of the Fianna Éireann, a heroic warrior band. Macpherson's Cuchullin, "general or Chief of the Irish tribes," is Cú Chulainn, the hero of an entirely different Cycle in Irish mythology.

We decided to tackle this from a networks point of view and compare Ossian to The *Illiad, Odysey* as well as *Acallam na Senórach*. The *Iliad* dates from the final year of the Trojan War (the eighth century B.C.) and tells of a dispute between Agamemnon, leader of the Greeks, and their hero Achilles. The *Odyssey* describes the journey home of Odysseus to his wife Penelope after the fall of Troy. The Finn Cycle of Irish mythology recounts the adventures of Fionn mac Cumhaill and the warrior band, the Fianna with *Acallam na Senórach* its most important source.

We used recent versions and translations: Gaskill's version of *Ossian*; Rieu's translation of the *Iliad* [Rieu, 2003]; Shewring's translation for the *Odyssey* [Shewring, 1980], and a recent translation of *Acallam na Senórach* [Dooley &

Roe, 1999]. We also examined *The Fianna*, Part II of *Lady Gregory's Complete Irish Mythology* [Gregory, 1904]. The latter is certainly derived from the Finn Cycle and therefore any attempt at structure comparison of two texts should place it close to *Acallam na Senórach*. We will see that our method succeeds in passing this test.

Our objective was to compare the network structures in *Ossian*, the *Iliad*, *Odyssey*, *Acallam na Senórach* and Lady Gregory's text. Many quantitative measures of network properties have been developed since network science emerged more than twenty years ago. We have measured many network properties, and the results have been published by [Yose et al]. but the essential information for our purposes is contained primarily in one feature: the degree distribution. To understand this, we must firstly construct the network.

To do this, we first carefully read the text, taking note of individual characters and meticulously recording the interactions between them. We do not require details such as the personalities or complexities of individual characters. Instead, we strip them down to their essentials and represent them by zero-dimensional nodes or points, because the set of interrelationships between characters, and not individual characters, primarily concerns us.

We found 325 characters (which are identified as nodes in the network) in Gaskill's version of the *Ossianic* texts: enough to perform a meaningful analysis. To set up the network, a reader has to decide on the nature of their relationship: if pairs of characters are related; if they speak directly to each another or speak about one another; of if they are physically co-present and it is clear that they know each other. In these cases, the nodes that represent them are joined by a link (also called an "edge"). If not, no link/edge is inserted. The nodes and the edges together form the *network*. The entire Ossianic network is displayed in Figure A.1

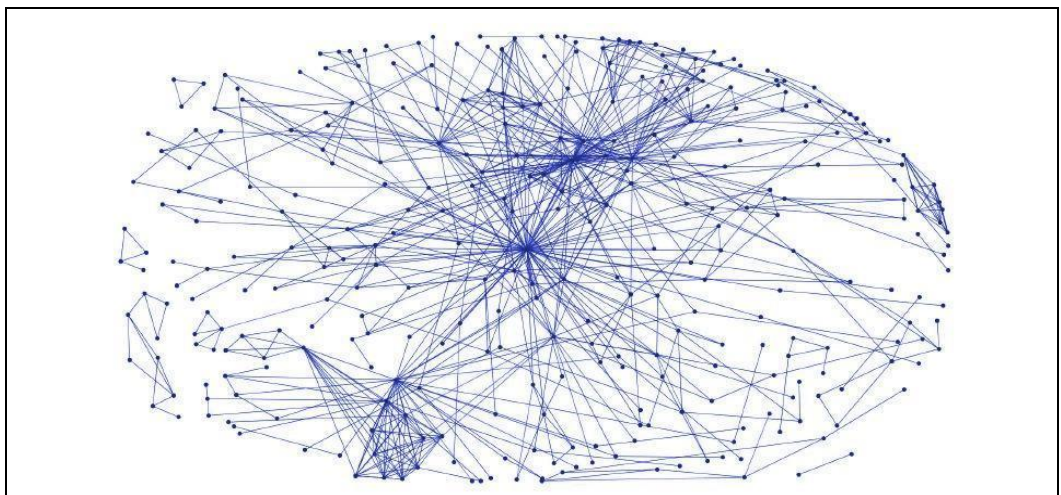

**Figure A1:** The entire network of 748 relationships between 325 characters depicted in Gaskill's version of Macpherson's Ossian.

To understand how the edges are distributed across the network, we offer Figure A2 as an example of a simple complex network. It is simple because it is small and easy to understand and complex as its topology is non-trivial. It is complex because it is not structured like a rectangular grid or a random graph.

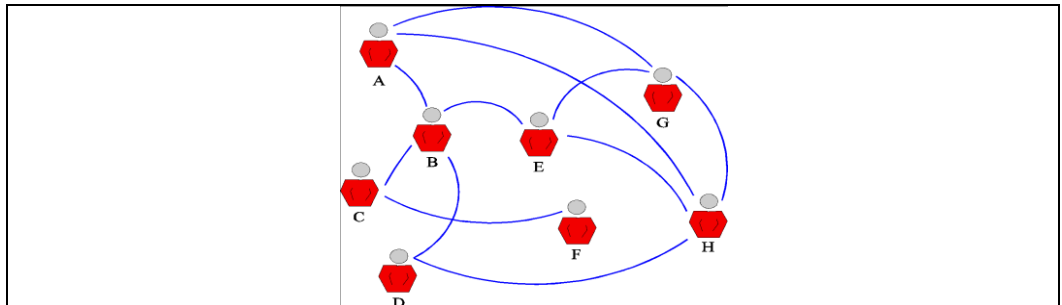

**Figure A2:** Networks comprise both nodes and links which represent relationships between them. In network theory, we are interested in statistically capturing how these relationships are distributed.

One of the most basic but essential network measures is the so-called *degree* of each node. This is the number of links that come from that node. We denote the degree of an individual node generically by *k* and, if we identify the particular node in question, we insert a subscript to label it. For example, in Figure A2, character A is linked with three other nodes (namely B, G and H). We say the degree *k* of node A is three or $k_A=3$. Different nodes have different numbers of edges attached to them and we are interested in how properties like this are spread over the network, from node to node. The proportions of nodes that have specific degree values are referred to as the degree distribution. We denoted the degree of node A in our simple network of Figure A2 by $k_A$, and we use analogous notation for other nodes. Counting the links of each node we find $k_A = 3$, $k_B = 4$, $k_C = 2$, $k_D = 2$, $k_E = 3$, $k_F = 1$, $k_G = 3$ and $k_H = 4$. We see that one node has degree one (*k*=1), two have degree *k*=2, three have *k*=3 and two have *k*=4. This set of numbers (how the degrees are distributed amongst the nodes) is called the degree distribution. We find it useful to look instead at the complementary cumulative degree distributions *P(k)*. This is the probability that the degree of a node is greater than or equal to the value *k*. In our case, five of the 8 nodes, for example, have degree greater or equal to 3, so we write $P(3) = 5/8$. Doing the same calculation for every degree, we find $P(1) = 8/8 = 1$. $P(2) = 7/8$. $P(4) = 2/8$ and $P(5) = 0$. Plotting these numbers on a graph gives us a visual representation of the spread of connectivity across the network.

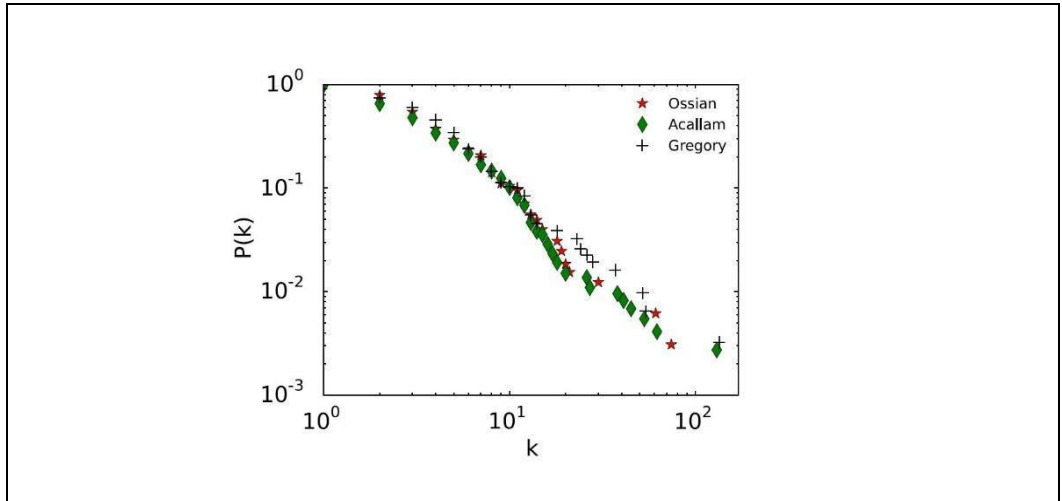

**Figure A3:** The complementary cumulative degree distributions of the full networks indicate that the society depicted in Ossian closely resembles those of the Irish *Acallam na Senórach* and Lady Gregory's text (denoted here by "Gregory").

In a larger network such as that we have generated for *Ossian* and the other narratives in this article, we prefer to present the complementary cumulative degree distributions *P(k)* rather than the degree distribution itself because the former reduces the noisiness in the plot. We present the complementary cumulative degree distributions for *Ossian* and the comparative texts in Figures A3 and A4. These figures represent the main result of this appendix. In Figure A3 we gather the cumulative degree distributions for *Acallam na Senórach* and Lady Gregory's text (732 and 301 characters, respectfully) alongside that of *Ossian*. The similarities are remarkable. They show that, at least in terms of the degree distributions, the network embedded in *Ossian* is very similar to those of the Irish texts. Moreover, the network in Lady Gregory's text matches that of *Acallam.* This is as it should be if the former is derived from the latter. Thus we can conclude that *Ossian* is as similar to and maybe even derived from *Acallam na Senórach* as is Lady Gregory's text.

We turn our attention to the Classics in Figure A4 where we plot the cumulative degree distribution functions for the Iliad and Odyssey (694 and 311 charcters) alongside that or Ossian. The dissimilarities are remarkable. Although they appear similar for small values of degree *k*, this is actually meaningless – all cumulative degree distribution functions start at the same place because the degree of all nodes is greater than or equal to zero for any connected network (i.e., all networks start with $P(k) - 1$ for $k = 0$). We can conclude that Ossian is as dissimilar to Homer's classics.

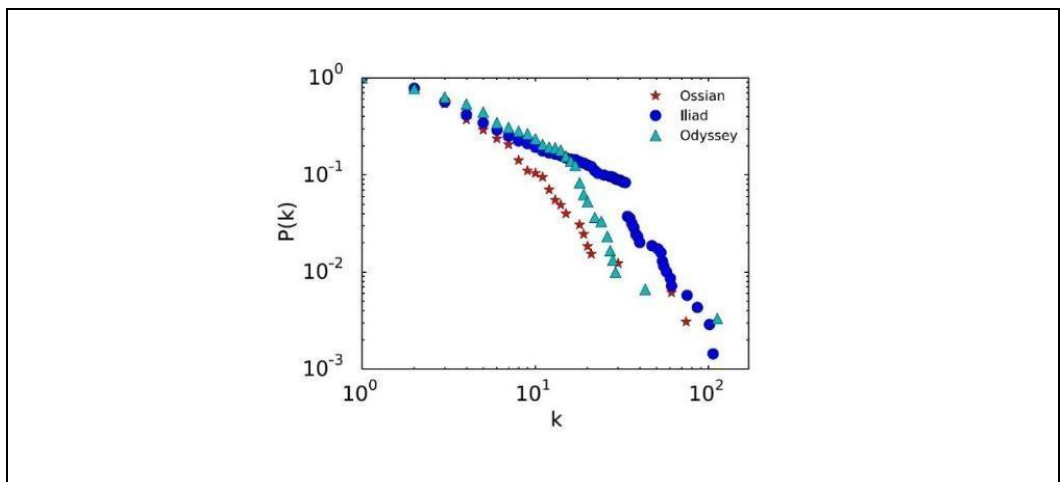

**Figure A4:** The complementary cumulative degree distributions of the full networks indicate that the society depicted in *Ossian* does not resemble those of either the *Iliad* or *Odyssey*.

Visual inspection of these plots delivers the main message of this appendix: even though Macpherson and his supporters sought to distance Ossian from Irish mythology, the networks unconsciously embedded in the two narratives are remarkably similar. Also, even though Macpherson and his supporters sought to align *Ossian* with Homeric epics, the networks inadvertently embedded in the two narratives are clearly different. These suggest that the Irish scholars were right and Ossian may have been misappropriated from Irish sources.

**Appendix B: The rise of Sinann**

In this appendix we record some of the letters in support of Sinann published in the pages of the *Westmeath Independent* and *Irish Post* newspapers in 2019 and 2020. We also record the petition signed by 700 people and the only written statement the council gave in support of the Mask. We document too the original poem "Ballad of Athlone" and the version adapted by Sarah Bailey for the Sinann campaign.

Appendix B1: Letters including the open letter

Ralph Kenna's first letter drew public attention to Athlone's incoming statue. Prior to this there were no objections.

**Letter 1: 25 May 2019**
**Call for rethink on Church Street sculpture**

Sir,

Many consider that art should provoke and that ensuing debates are welcome. It is in this spirit that I request that you publish my opinion on the new Church Street sculpture.

I greatly admire your writer Gearoid O'Brien; his Miscellany and his contributions to local history are of exceptional and enduring value. Therefore I can only assume that the positivity of his review of the Church Street sculpture (4 May edition) is driven by kindness and

None of the symbolism, or any of its creators, has anything specific to do with Athlone. At best the proposed monument represents deference to Dublin and maybe to Limerick and maybe even London. The idea of using neo-classical river heads for the keystones stems from Somerset House in London. This was designed by Sir William Chambers (master of the Custom House architect) and its adornments denote the products of the principal rivers of England.

Although I agree with Gearoid's comment "For far too long Athlone has had very little in terms of quality pieces of public sculpture", there are appropriate examples we can be proud of. The new addition does not belong to this category and instead reinforces colonial attempts to eclipse local heritage.

| | |
|---|---|
| courtesy rather than critique of the object itself. This communication aims to address some important omissions.<br><br>The "river god of the Shannon" as presented in this sculpture does not in any way "hark back to Irish mythology" or, indeed, Athlone heritage. Our river was named after Sionann – the granddaughter of Lir (of the "Children of Lir" fame). This unfortunate sculpture therefore represents misappropriation of gender and aggrieves women worldwide whose names (Shannon) derive from our river deity.<br><br>But the effrontery (to my mind at least) is even more than that. As Gearoid's item so ably demonstrates, the notion of a (male) "river god" of the Shannon was entirely concocted in the 18th century for the Custom House in Dublin.<br><br>The river-effigies on the Custom House represent commerce and profit – hence their appearance on our national currency – and have nothing to do with our mythology. The prime position of the custom-House ornamentation is held by the crown of England, perched superior to the harp of Ireland which is surrounded by the lion and unicorn – symbols of governance in a time of subjugation. | I understand that three shortlisted pieces were exhibited for public viewing in the Civic Centre.<br><br>But either the iconographic and factual information I give above were not to hand or not adequately promulgated to enable Athlonians to deliver fully informed opinions in the narrow timeframe available. I therefore suggest a return to the drawing board to add to the sculptures that actually represent our town and that we actually can be proud of.<br><br>Ralph Kenna,<br>St Brigid's Terrace, Athlone<br>and<br>Murray Road, Rugby,<br>England.<br><br>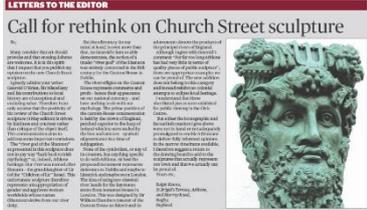 |

The open letter was signed by 127 people when it was submitted to the *Westmeath Independent*. Signatories were from all quarters of life and all professions. Amongst the notable signatories were Professor Tomas Ising, son or Dr Ernst Ising whose work was pivotal to modern theories of phase transitions and critical phenomena. The first citation in Ising's thesis is to Richard Kirwan, whose contribution to the research underlying this campaign is discussed in the main body of the text.

| | |
|---|---|
| **Letter 2: Open Letter 27 July 2019**<br>**Planned Athlone artwork a 'misappropriation of our heritage'**<br><br>Dear Editor<br><br>We refer to the new Church Street sculpture intended to "address Athlone town's urban context, the location of the artwork close to the River Shannon and the Old Rail Trail cycleway, as well as the heritage, memory and environment of the town." As you know, the piece selected by Westmeath County Council is adapted from the Custom House in Dublin where it is claimed to represent a Shannon river deity.<br><br>On behalf of hundreds of people from Athlone, people who have settled in Athlone, and Athlone's national and international diaspora, and on behalf of concerned people nationally and the world over, we object to the intended statue. It represents misappropriation of Athlone's and Ireland's heritage, an affront to Mná na hÉireann and to people who value parity.<br><br>The Custom House iconography is neo-classical and is not at all connected with Irish heritage. It claims a vacuous male god to represent our river while in Irish mythology the river is female. The name Shannon stems from Sínann (Sinann, Sionann, Sionnainn, Sinand, Sineng, etc.) and may derive<br>from "Old Honoured One". It is personified as a goddess – it is female, not male, and millennia older than the | You claim "the reverse of the piece depicts a complex weave of circuitry. Visually this refers to the increasing 'connectedness' of our world and the myriad of connections within Athlone, both geographical and infrastructural (road, rail, river)." Athlonians consider the town to have value in its own right, not merely as a conduit from between one location and another. And we find it inappropriate to represent our town on the reverse – or backside - of any statue. The council have indicated that the piece may ultimately end up somewhere less conspicuous. But even that is unacceptable - if it's wrong, it's wrong.<br><br>The public were allowed to view three shortlisted models for two weeks (although their observations would "not be counted"). But the above iconographic information was not available for our people to deliver fully informed opinions. The council's claim that "the process for the selection of the art piece was a detailed and considered process", "conducted to a standard that adheres to policy and procedures" does not refute any of our points. Indeed, since the process delivered an inappropriate piece, rejected by the people in the light of new information, then that very process is flawed and<br>should be annulled.<br><br>Some of the above concerns have been delivered to the council by the first signatory of this letter. The local authority's response, claiming to have been "fully cognizant", that the decision was "a fully informed one", |

Custom House. The Metrical Dindshenchas (our native recount of the origins of place-names and traditions, committed to writing over a thousand years ago) relates how Sionann drowned at Connla's well in her pursuit of wisdom. "Rivers in Ireland … were envisaged as divine figures to whom were attributed the gift of poetical inspiration, mystical wisdom and all-encompassing knowledge."3 Ours is a rich and ancient mythology, one we should be proud of and take inspiration from.

In contrast, the Custom House in Dublin is devoid of cultural meaning. It was designed in the 18th century by James Gandon whose world was one of privileged nobility. Beneath that world lay "the impoverished underclass … still struggling for their basic civil rights."

Elitists like John Beresford, the first commissioner of revenue in Ireland who conceived the construct, insisted that they "should be kept down by a policy of unyielding repression." When Napper Tandy (co-founder of the United Irish movement of Catholic, Protestant and Dissenter) et al protested, Beresford instructed Gandon to carry on and to "laugh at the folly of the people."

Thus, and as pointed out previously in public, the use of Custom House motifs for Athlone is a discordant move to say the least. The Custom house ornamentation positions Shannon mythology away from Irish sources and closer to the classics. This echoes the Ossian debacle 250 odd years ago – an attempt to supplant Ireland's mythological heroes to position a concocted narrative away from Irish mythological sources and close to the classics. It represents arrogant and brutal misappropriation of Irish mythology. The Custom House theme also misappropriates gender; although our town has no female iconography, you, our Council, deem it appropriate to commission a concocted male figure instead of the authentic female one which is steeped in our heritage and story.

and that "the process is underway and cannot be reversed", more resembles Beresford's instructions to "laugh at the folly of the people" than it does Sionann's pursuit of wisdom.

In her article 'The Fate of Sinann', Kiltoom-born Celtic scholar Maud Joynt (1868-1940) discusses how the "the original legend perhaps foreshadowed the dangers" of insufficient expertise. As elucidated by Professor of Irish Studies Noémie Beck, "knowledge was believed to be perilous when not handled correctly".5 Thus, in Sionann's story it appears that our ancestors warned of the danger of knowledge without context – the very pit that the council appears to have lead us into.

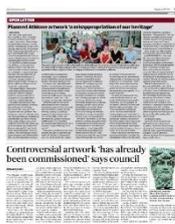

We call for these words to be heeded and this folly to be halted. The Council, and our representatives, still have a chance: Follow our much loved Sionann and embrace the new knowledge and context you previously did not have. In doing so, they will be remembered as enlightened representatives who listened to the people of the present and sages of our past. Or follow Berresford and associates, carry on without regard for our heritage, and be remembered in an altogether different light.

The choice is yours.

Signed: Professor Ralph Kenna, Orla Donnelly, Actuary and Community Volunteer, Athlone, Fiona Lynam and over 100 (names with editor).

Kenna's third letter to the *Westmeath Independent* (below) was published alongside Caroline Mannion's letter.

**Letter 3: 14 September 2019**
**Public artwork selection was flawed from the outset**

Sir,

I address your informative report "No u-turn on controversial Athlone sculpture" (*Westmeath Independent*, September 7).

As is now widely known, the 11-ft neoclassical statue intended for Athlone takes its iconography from 18th-century colonial misappropriation of Ireland's heritage by replacing an authentic female river deity by a male one concocted for commercial representation of the Custom House in Dublin.

Despite warnings published in your newspaper over the past three months, the Council appears intent on echoing instructions by the Custom House's commissioner to "laugh at the folly of the people" by ignoring new iconographic information missed by them at the time the piece was selected. Much of the new information is based on my own research into cultural misappropriation from the same period as the Custom House.

The public has embraced this new information, and people initially in favour of this statue are now outraged that they were not informed as to its

Instead of blaming the public and academics the Council should accept that the process was flawed from the outset. To evidence this, I cite Cllr Keena's comment "art is art" and "some people won't like it from a visual point of view".

Cllr Keena should by now be fully aware that the protestations have nothing to do with "visual" aspects and, if he is not aware, he should read the open letter (July 27) signed by folklorists, medievalists, physicists, mathematicians and now informed people from a vast diversity of lay backgrounds. Visual art is one thing but representational art is another. This piece was never primarily intended as an addition to the aesthetics of the town - it is to represent the town itself, its people and its heritage.

Comments that reveal lack of knowledge of iconography and symbolism undermine credibility and negate the Council's procedural argument. This statue portrays our river, Athlone and Athlonians in a very bad light. It is an 11ft monument to the colonial suppressors of our ancestors, to the Council's ignorance, and the arrogance of its intent on ignoring academic information and informed public opinion. As such, it does not represent a now *informed* Athlone.

Yours etc,

| | |
|---|---|
| symbolism. Withholding information is deceitful, but I'm sure this was not the Council's intent – they simply did not have sufficient expertise.<br><br>At no stage in the discourse did the Council rebuff any of the new academic, symbolic or iconographic objections taken up by the public. Instead the Council's sole justification to proceed is to repeat "process is process". In doing so they apportion the blame to two quarters: 1)The people who supported for the intended piece; and 2) Academics in general, and me in particular, because intervention came late.<br><br>Both are disingenuous: (1) because the people did not receive the very information they trusted the Council to give them and (2) because the public consultation process was too narrow to reach the academic community expeditiously.<br><br>Moreover, the council stated that votes would "not count." If numbers are to count retrospectively, the petition (delivered to the Council on 29.08.2019) and near universality of opposition negates the Council's post-hoc argument.<br><br>A credible (if troubling) reason for Council's stonewalling is their statement that "changing course at this stage would undermine the credibility of the councilors." This is akin to Cpt Smith's refusal to change the course of the Titanic despite warnings of what lay ahead. A little embarrassment now will prevent enormous and enduring embarrassment in future. | Ralph Kenna<br><br>PS: On 02 October I will deliver a lecture in the Sheraton Hotel hosted by the Old Athlone Society. This lecture is open to the public and will address the academic research into misappropriation of Ireland's heritage which triggered this controversy. It will explain why this statue is wrong. As pointed out in the open letter mentioned above, Sionann's own story illustrates that insufficient knowledge led to her demise. If councillors wish to avoid a similar fate for their reputations, I advise them to attend my lecture.<br><br>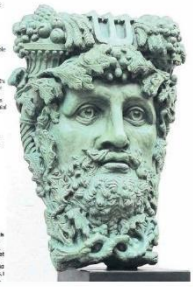 |

Caroline Manion's letter was published alongside Kenna's third.

| | |
|---|---|
| **Letter 4: 14 September 2019**<br><br>**Council is 'airbrushing' women from history in statue row**<br><br>Sir, Irish women were airbrushed from history until relatively recently. Luckily with the centenary of the 1916 Rising women who played such an important part in the fight for Irish freedom were recognised, including two local women the Elliot sisters, who were rightly commemorated in their home town. However the local council have decided to airbrush the Celtic goddess of the Sionnan and replace her with a male river god. It seems a long time since Mna na h-Eireann had to fight for their place in Irish society but the local council doesn't seem to mind changing a female entity into male one like so many have done in story telling in the past. The council's insistence that the public picked the winning sculpture doesn't hold up. The first time the public saw the three shortlisted sculptures was in the *Westmeath Independent* and there was no explanation as to the meaning of any of them.<br><br>I don't believe that if the people of Athlone were fully informed about the origins of the name of the River Shannon and were told that we were having a male river god instead to represent our great river they would have voted for it. Once that mask was aligned with the Shannon and not just a random head it became wrong. | It doesn't matter if it's a great work of art, it could be Michelangelo's Pieta but if it was representing the River Shannon it still would be wrong. Art isn't just art.<br><br>I find it very patronising to be told that the people of Athlone often don't like things but when they are done then they like them as our Minister recently said in relation to the Luan Gallery and even if people do get to like things they hadn't previously I don't think they believe a female goddess should be transgendered by the local council into a male god.<br>Yours sincerely<br>Caroline Mannion,<br>Bloomfield Drive,<br>Athlone |

Throughout the entire campaign, only one letter (below) was published that voiced support for the Mask of the Shannon.

| | |
|---|---|
| **Letter 5: 14 September 2019**<br>**Statue is valid work of art for Athlone** | not so distant past is just as authentic and valid as anything inspired by the ancient mythological past. |

Sir,

Did anybody ever think of going down to the Shannon and asking the river, "are you a man or a woman?" No? I lived and worked in Athlone for three years and n all that time I never once heard of the goddess Sionann and I dare sy not many other people in town have heard tell of her either. Wouldn't resurrecting the old pagan goddess now not be an act of cultural appropriation par excellence? I believe that a work of art which is in conversation with the Perhaps if people weren't stuck in their ivory towers they would be aware of the consultation process which preceded the commissioning of what appears to be a very find piece of art. I trust said sculpture shall soon be erected by the banks of the lordly River Shannon.

Kenna's fourth letter (below) sarcastically addresses the Councils repeated suggestions of a second statue in support of Sinann. The Campaign rejected his because "a right does not negate a wrong".

Letter 6: 16 November 2019
**New sculpture 'should be outside council office'**

Sir

As the 11ft "Mask of the Shannon" relentlessly marches its way to Athlone, I have been racking my brain trying to find a way to mollify the imposture.

The Council's suggestion (*Westmeath Independent* 19 Oct) of a second sculpture does not of course reverse the offence of the first, as a right does not negate a wrong.

I have a suggestion that goes some way to address the symbolic affront while matching the Council's insistence that they are "not for turning" in erecting their mask that they consider "looks nice."

Despite being invited, not a single councillor attended my public lecture (2 Oct) where I explained why the statue is fundamentally wrong, standing for colonialism, misogyny and misrepresenting Athlone.

I therefore emailed a link to a video of my talk but unfortunately no councillor responded.

Since the Council ignore direct communications from me, I hope publishing this letter openly will be a more effective means to propose a way forward.

Instead of erecting the mask at Custume Place as a representation of Athlone and Athlonians, erect it directly outside the Civic Centre as a representation of the Council and councillors themselves.

Moreover, if the "complex weave of circuitry" that is supposed to represent Athlone is oriented towards Church Street, the mask would be on the backside facing the council's offices.

This would match the usage of the same iconography on the reverse of the old Irish 50-pound notes in that it represents commerce instead of culture.

Additionally, Councillors would have the pleasure of viewing what they consider "nice" while we Athlonians would not have to suffer the indignity of a misogynistic mask misappropriating our mythology and heritage.

Thus all concerned would be happier and Custume Place can be freed for a future piece that will actually represent the town.

Yours erc,
Ralph Kenna
2 St Brigid's Tce.
Athlone.
PS: The link to a video of my Old Athlone Society lecture s here: https://www.youtube.com/watch?v=MDujKENPnQc

Over the course of the campaign, Kenna also published a letter in *The Irish Post* newspaper (below).

| 11 July 2020 **Controversy over Shannon statue** | Protests include a petition, so far signed by 650 people, peaceful street events, art, poetry and a Facebook forum. An |

Dear Editor

As some of your readers know, a new statue is on its way to the heart of Ireland and over the past year has been causing outrage locally, nationally and internationally - even before it is erected! Since the issue colonial statues is very much to the fore globally now, I thought I should update your readers as to how Athlone getting along.

It started last year when Westmeath Co. Council called for a sculpture to represent "the heritage, memory and environment of Athlone" and in particular the Shannon River on which the town stands. A 3.5m Neptune-like figure claiming to "hark back to Irish mythology" was selected by an anonymous panel as the "fierce and proud river god, laden with fruits of the river's basin."

However, in Irish mythology, the Shannon is named after a goddess – not a god! Her name is Sionann and she is the granddaughter of Lir (of "Children of Lir" fame). To overwrite her by an 11ft hairy male statue is considered by many as an affront to our ancient heritage, an affront to Mná na hÉireann, and an affront to egalitarians the world over.

It is also not specific to Athlone. The male "river god" of the Shannon was concocted in the 18th century for the Custom House in Dublin and now adorns inner pages of new Irish passports. The river-heads on that historic building represented commerce and profit and the export of Ireland's resources to the British Empire. The prime position of the Custom-House ornamentation is still held by the crown of England, perched superior to the harp of Ireland which is surrounded by the lion and unicorn – symbols of governance in a time of subjugation.

open letter calling for the statue to be revoked was signed by 150 people of diverse backgrounds including mythologists and mathematicians, folklorists and feminists, counsellors and carers.

The statue is not yet up, and Westmeath Co. Council has a choice. They can follow the Custom House architect whose instructions were to "prevent all opposition and laugh at the folly of the people." Or they can acknowledge Sionann and what she represents. The question is whether colonial iconography that overwrites Mná na hÉireann is apt representation of Athlone's and Ireland's identity into the future.

Regards
Ralph Kenna

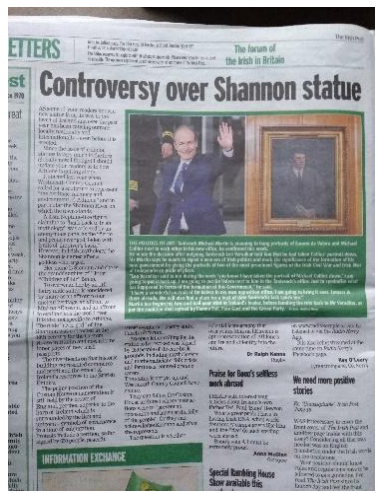

Kenna's fifth letter in the *Westmeath Independent* addressed comments by the artist Rory Breslin as does Caroline Mannion's second letter.

**01 November 2020**
**We must reclaim or heritage in statue debate**

Sir,

In his interview about the sculpture intended for Athlone, Rory Breslin argues it has "a strong Irish theme" and is "really not sure how anybody could argue" that it's based on British or colonial symbolism.

Allow me to explain: Representation of the "heritage, memory and environment of Athlone" and "in particular the River Shannon"[69] in neo-classical Neptunesque form overwrites Sinann, the authentic cultural identity of our river. Overwriting culture and identity are precisely the aims of colonial assertions of supremacy. This is most blatantly achieved in the Custom House by positioning the British Crown superior to the Irish Harp and surrounding it by the lion and unicorn. It is also achieved by the inscription "1690" on the Custom-House figure meant to represent the Boyne. The Boyne and the Shannon have interrelated narratives in Irish mythology and 1690 was also the year of the first Williamite siege of Athlone.

Referring to 1791 as the date the Custom House was completed, Breslin states "Given that there was an Irish parliament at this time, the building was a demonstration of the aspirations of the Irishmen who were responsible for running the country." True – but the aspirations of that parliament were sectarian! They ran counter to the aspirations of the United Irishmen who sought to "unite Protestant, Catholic and Dissenter under the common name of Irishmen" (sic) and "subvert the tyranny of our execrable government". Custom House architect James Gandon fled back to London while the United Irish rebellion (1798) was crushed with unsurpassed brutality, followed by plunder rape and murder.
The currency symbolism that Breslin refers to in his interview was precisely that – currency! The keystones on the Custom House were designed to represent extraction of wealth of Ireland for taxation and export to the British Empire. Their monitory symbolism was adapted by the Free State government as standing for commerce not culture; profit not people; money, not mythology.

---

[69] Source: "Sculpture to reflect town centre facelift, Westmeath Independent, 26th July, 2018.

Finally, I remind the issue of gender. It is wholly inappropriate to import a concocted colonial colossus from Dublin to represent a town with no secular female iconography and it is downright outrageous to overwrite an authentic female figure (Sinann) in doing so.

We, as Athlonians, clearly need to revisit our past and reclaim our heritage to design our future. We simply cannot look up to a figure that represents money, misogyny and misappropriation as our inspiration. Over the next few weeks, an arts competition will be launched to promote the authentic river deity Sinann. Over €4,000 in prizes will be made available and the competition will be announced in *The Irish Post* and *Westmeath Independent*. I invite all Athlonians to participate.

Regards
Ralph Kenna
St.Brigid's Tce

## 01 2020
## It's not about whether the Mask is good art

Sir,

By trying to defend his work Mask of the Shannon Rory Breslin managed to prove the opposite.

His claim that since the Custom House was built in 1791 and Ireland had a parliament then, means that there wasn't British symbolism involved in the design of the original masks.

His definition of an Irish parliament that was running the country is a lot different from what most Irish people see as an Irish parliament. This was a protestant only parliament since catholics were excluded from power. Ireland was stripped of its resources and many people died in a famine long before the more well known famine in the 1840s. After the plantations many of the English were absentee landlords with the farm work being done by the Irish poor on land that was formerly theirs.

Even after Henry Grattan had achieved some independence for the parliament he was still a member of the Anglo-Irish elite and still insisted that Ireland remain linked to Britain by a common crown. Eventually Catholics were given equality in certain areas. Irish and gaelic culture was denigrated and our myths and history were treated as inferior to the British fondness for Greek inspired culture. That is why the river masks are based on classic Greek myths rather than our own Irish mythology.

The British crown sits upon the original masks on the Custom house. Our rivers were named after female deities as we all know by now. We are not arguing whether the masks are good art and I'm sure the originals were very well crafted but I presume the statue of the slave trader recently toppled in Bristol was a fine example of art too.

However, like the Mask of the Shannon it might have been admired when it was erected but now people know better. It's only four years since we all celebrated the centenary of the 1916 Rising. All the politicians from the different parties on the council were proud to take part and celebrate the start of our independence. I thought in 2020 we would all be proud of our Irish culture and not still feel that if something came from some other country it must be superior.

Yours, etc.
Caroline Mannion
Bloomfield Drive
Athlone

Two years on, in 2022, the issue still had not gone away with Cllr O'Rourke mentioning in a comment that he hopes the next artwork won't be so controversial. This prompts a response by Kenna.

| | |
|---|---|
| 03. 12. 2020<br>**The Per Cent For Arts scheme**<br>Sir,<br><br>I read with interest your item (19 November) "Five new sculptures set to be installed on greenway".<br><br>I was bemused by Cllr O'Rourke's comment "I only hope we don't see the same controversy as the last Per Cent for Arts scheme project we took on in this council." As your journalist writes, he was, of course, "referring to the huge controversy that surrounded the installation of the so-called 'Mask of the Shannon' project."<br><br>Mr O'Rourke's concern targets the controversy rather than the offence that caused it. Calling county councils to account is an integral feature of democracy and such a comment cannot go unchallenged.<br><br>To remind your readers, the "Mask of the Shannon" is a colonial "river-god" from the 18th century that overwrites our beloved "goddess" from Irish mythology.<br><br>Her name is Sinann and the purpose of her replacement was to overwrite Ireland's national identity and that of the River Shannon.<br><br>The "huge controversy" to which Mr 'Rourke refers was the usage of this colonial iconography in Athlone to represent the town's heritage and relationship to the Shannon despite its colonial origins and despite it having nothing to do with the town.<br><br>But the controversy was much more than that - it was also about masking over female mythological symbolism by a male motif, an offense that is additionally aggrieving in a town that has no secular female iconography. | However artistically admirable the five new statues may be, they do not appear to address that imbalance; Athlone continues to have no female iconography!<br><br>In case the council has completely forgotten, let me remind that half our population is female. Surely, we can find a few Athlone women worthy of a statue.<br><br>If Mr O'Rourke and the council can't think of any, I advocate for Sinann herself or another figure from Ireland's mythology.<br><br>As you reported in your issue of 03 Sept 2022, we are running a €10,000 art competition centred on such figures. With two hundred entries received so far, including many from local artists, there is plenty that can inspire.<br><br>Results will be announced in the new year and will hopefully enter the minds of Westmeath County Council the next time the Per Cent for Arts scheme presents an opportunity for new statues.<br><br>As a bonus, when Athlone has properly gender-balanced iconography Mr O'Rourke may have fewer controversies to fret about.<br><br>Regards<br><br>Ralph Kenna<br>St Brigid's Tce<br>Athlone<br>Tel: +44 7557425143 |

## Appendix B2: Petition and cover letter

The petition was initiated and submitted by Fiona Lynam.

| | |
|---|---|
| **From:** Fiona Lynam<br>**Sent:** Wednesday 28 August 2019 15:53<br>**To:** Fiona Lynam <fiona.lynam@socialdemocrats.ie><br>**Subject:** Proposed Statue for Athlone<br><br>Dear<br><br>We attach for your attention a public petition that has generated 586 signatures to date via Social Media platforms. The petition was raised in order to gauge public interest in points raised by noted mythology and physics professor Ralph Kenna regarding the appropriateness of the chosen Rory Breslin designed statue as an icon representing Athlone.<br><br>By engaging with the people directly it is quite clear that the public at large were not fully aware of the meaning of the poposed Greek God Head statue as chosen by the council in early 2019. It was not until Prof Kenna started his campaign of public education via local radio and newspapaers that it became clear that this Greek God Head represents not only a celebration of British colonial rule but is also a misappropriation of the story, gender and heritage of the Goddess of the River - Sionnan.<br><br>Whilst we all appreciate a set process was engaged with when this statue was chosen by the council and their panel of experts it is a widely held belief that this process was flawed. The timing of the public consultation was poor taking place as it did in January at a time when people are less likely to be interested in engaging with public consultations due to bad weather, dark evenings and lack of resources. Whilst the public were invited to cast votes for their preferred statue, public opinions were not part of the consultation process. Whilst we understand this approach we feel the issue that was missed in regard to this statue is that the image itself has a history. It is not an imagined design from an artist. It is a direct copy of a carving that comes from the Custom House in Dublin which as a result has a history, a meaning and a message. That history, meaning and message deserved to be fully explored and most importantly explained to the public before it was selected.<br><br>Please note that this petition and campaign are not concerned with complaining about the artistic merit or otherwise of the chosen statue. We are all aware that art is subjective some people will like it, others will not. That's all granted. The issue that we are raising relates directly to what this image tells us about Athlone. How it represents the history of the Shannon and our town. We all want to be proud of it but as it stands it is a wholly inappropriate and deragotory representation of all that we should be proud of.<br><br>As such we are asking you to enage with us to find a means of salvaging this situation. We are calling on you to work with us to find a way of presenting a piece of public art that the whole town can be proud of or at the very least won't be ashamed of.<br><br>Is mise le meas,<br>Fiona Lynam<br>(on behalf of the Reclaiming Sionnan for Athlone Campaign) | Reclaiming Sionann for Athlone<br>FL<br>Campaign created by<br>Fiona Lynam<br><br>Chose an appropriate street installation for Athlone that celebrates our heritage, our people and our town<br>Why is this important?<br>The "river god" statue chosen for Athlone misrepresents our native heritage and our rich culture.<br>It would be an affront to our heritage and people to use colonial male object from Dublin to represent the Shannon and Athlone.<br>The mythological Goddess Sionann, granddaughter of Lir, is our mythological river deity – not a concocted neo-classical god.<br>Misappropriation of mythology and gender in a time of national subjugation is not acceptable as a modern representation of our town, nor is representation of the town on the backside of a statue.<br>The concept of celebrating our river and our heritage is a welcome one and we call on the Council to do so by recognising our heritage – not replacing it.<br><br>We call on Westmeath County Council to revoke its uninformed selection. |

Appendix B3: Council's written statement

At the meeting of 2 July 2019, the Council penned the following letter and requested it be read out at radio interviews etc. (Note the date on the letter is mistakenly 3 July.)

[Handwritten meeting note dated 30/10/19, transcription:]

> 30/10/19
>
> Following a meeting today between Orla Donnelly and Ralph Kenna and elected members Cllrs Frankie Keena, Aengus O'Rourke, John Dolan and Louise Heavin
>
> The elected members present stated that the selection of the new art piece for Church St was identified and selected through our County Council Public Arts Policy and due process was carried out.
>
> Public consultation was also carried out and we thank the public for engaging.
>
> The art piece has been commissioned to the artist and it is planned to be erected before the end of the year.
>
> The members present feel that some people will like it and some may dislike it from a visual point of view. They are happy with the piece as an appropriate art piece for the town.
>
> We invited Orla + Ralph to engage in the process for a planned new art piece (pending announcement by the OPW) for Athlone to represent the Shannon Goddess.

Appendix B4: Sarah Bailey poem and original Ballad of Athlone

The poem "A Ballad for Sinann, a Ballad for Athlone" was published on 10 October 2020. It is based on Aubrey de Vere's poem "The Ballad of Athlone".

| **The Ballad of Athlone** <br> **By Aubrey de Vere** | **A Ballad for Sinann, a Ballad for Athlone** <br> **By Sarah Bailey** |
|---|---|
| Does any man dream that a Gael can fear? <br> Of a thousand deeds let him learn but one! <br> The Shannon swept onwards broad and clear, <br> Between the leaguers and broad Athlone. | Does anyone dream that a Gael can't care? <br> Of a thousand wrongs let them learn this case! <br> The Shannon swept onward, broad and clear, <br> Between the Castle and Custume Place. |
| 'Break down the bridge!' - Six warriors rushed <br> Through the storm of shot and the storm of shell; <br> With late but certain victory flushed. <br> The grim Dutch gunners eyed them well. | "Halt the statue!"— the Wild Geese rushed <br>   Through the storm of conceit, the truth to tell: <br> With late, but certain embarrassment flushed, <br>   The grim Town Council eyed them well. |
| They wrench'd at the planks 'mid a hail of fire; <br> They fell in death, their work half done; <br> The bridge stood fast; and nigh and nigher <br> The foe swarmed darkly, densely on. | They wrenched at the arrogance mid a hail of ire; <br>   They wrote their letters, their work half done: <br> The Council stood fast, and nigh and nigher <br>   The statue marched darkly, densely on. |
| "Oh, who for Erin , will strike a stroke? <br> Who hurl yon planks where the waters roar? <br> Six warriors forth from their comrades broke, <br> And flung them upon that bridge once more. | "O, who for Sinann will strike a stroke? <br>   Who'll halt yon statue where the waters roar?" <br> 100 Athlonians from their slumber awoke, <br>   And penned the newspaper letter once more. |
| Again at the rocking planks they dashed; <br> And four dropped dead, and two remained; <br> The huge beams groaned, and the arch down-crashed - <br> Two stalwart swimmers the margin gained. | With myths and maths and historical fact; <br>   They signed the petition, and danced for her name: <br> The Council groaned, and sealed their contract; <br>   While stalwart Athlonians their heritage gained. |
| St. Ruth in his stirrups stood up, and cried, <br> "I have seen no deed like that in France !" <br> With a toss of his head, Sarsfield replied, <br> "They had luck, the dogs!'Twas a merry chance! | The world in amazement stood up, and cried, <br>   "We're pulling down statues like that in France!" <br> With a toss of their heads Sarsfield Square replied, <br>   "They misinformed us, the dogs! They gave us no chance!" |
| O many a year, upon Shannon 's side, <br> They sang upon moor and they sang upon heath, | O, many a year upon Shannon's side |
| Of the twain that breasted that raging tide, And the ten that shook bloody hands with Death! | They will sing upon moor and they'll sing upon heath Of the 600 people who signed with pride And their goddess eternal at swim beneath. |
| | |
| **28 November 2020, 19 December, 16 Janury 2021,** | 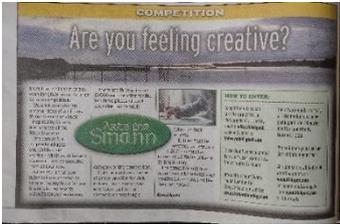 |

## Appendix C: Notes on Ó Brolcháin Carmody's translation, O'Curry's version and accompanying notes by Ó Brolcháin Carmody

In this appendix, and for completeness, we supply (C1) notes made by Ó Brolcháin Carmody on her translation (Section III.B). We also supply (C2) O'Curry's version with critical notes by Ó Brolcháin Carmody.

Appendix C1: Notes on names and places that occur in the dindshenchas by Ó Brolcháin Carmody:

*Condla/Connla:* This is still a male personal name in Ireland. The poem indicates the fittingness of this well-being under the sea, implying a general understanding that this is Connla's realm. Its derivation seems to be from the verb con-dáili, meaning to share (in) or meet up. There is possibly another sense (under cunnla in eDIL) meaning "modesty" or "propriety". This cluster of meanings is also found in the alternative name for Connla's Well of Lind Mná Féile – see below.

*Crimthann:* This name is explained in the medieval glossaries as sinnnach, "fox"; but it is worth bearing in mind the glossators (O'Cléirigh in this case) may already have the association between Sinann and Crimthann from some version of the Dinshenchas story. It may, however, be related to another word, crem / crim, "dog leek, wild garlic".
The text "Cóir Anmann", The Fitness of Names, functions quite like a glossary, as the author(s) often has to cast far afield to understand some archaic point of language which he doesn't understand. For example, at least one entry "analysing" a name beginning with Eo- (which we now know either to be *ío, "yew", éo, "salmon" or part of e[o]ch, "horse") translates this element as the Greek Eu-, "good". Fortunately for us, the text also contains scraps of story that we can match up with other sources. Here is an entry concerning someone called Crimthann:

> *Crimthan Nía Nár . Níadh .i. trén .i. trénfear Naíre .i. Nár thúathach a sídhibh, ben Chrimthain . Is sidhe rug Crimthan lé a n-echtra n-ordhairc a Dún Chrimthain a n-Édur*

> *Crimthan Nía-Náre: nía i.e. strong i.e. that is Nár's champion. Nár, a witch from the sídh mounds, was Crimthan's wife. It is she that took Crimthan with her on the famous adventure from Dún Crimthain on Howth.*

Cóir Anmann, edited by Whitley Stoes can be found here: https://celt.ucc.ie/published/G106500C/index.html

*Cú Núadat:* Literally, "The Hound of Nuada". I have been unable to find any other reference to a person called Cú Nuadat, but there are many to Mog Nuadat. Mog or Mug means "slave / servant", and I don't think it's a huge stretch to suggest that the "Servant of Nuada" and the "Hound of Nuada" might refer to the same person, or at least to people in a similar position.

Nuada himself is the mythological king of the Túatha Dé Danann, who lost his arm at the first Battle of Moytura, and losing the kingship because of this blemish. He was furnished with a silver arm, and thence became known as Nuada Airgetlam ("Silver-Arm"). There is little doubt that Nuada (*Nódo in older texts) is the same name as Nodons, the British personage who gives us the Land of Nod et al. We discuss Núada and the Battle of Moytura in detail in Series 2 [*Story Archaeology*].

*Imbas*: Imbas is a gift or technique central to the poets of ancient Ireland. Its literal meaning is imb-fhes, "great knowledge". Gwynn consistently translates this as "magical lore" in the Sinann poems, but it is perhaps best understood as "inspiration". In many texts, it is associated with rivers and wells, especially the wells of Segais and Connla, the mythical sources of the Shannon and the Boyne. O'Davoran's Glossary includes a description of imbas greine, imspiration from the river Graney, by which a poet gains (i.e., creates) a poem. There is also a poetic technique called sretha imbas, stream of inspiration, where every word in a line of poetry alliterates; and a metre called imbas forosna.
Imbas on its own is often taken to stand for imbas forosna, "great knowledge which illuminates". As well as the name of a metre, it is also described as a technique associated with the highest grades of poets.
For more on this, see the article "Imbas: Poetry, Knowledge and Inspiration."

*Lind Mná Féile:* Literally "The Pool of the Generous Woman". Féile is a virtue singularly associated with women, and is therefore often translated as "chastity" or "modesty". It seems to me to have a root meaning of "generosity" or "hospitality", and this seems to have developed into a concept of propriety. It is also interesting to note that female genitalia are sometimes referred to as Féile!

*Lodán:* Sinann's father, is mentioned in ¶ 17 of Acallamh na Senorach ("The Colloquy of the Ancients"), as the son of Lír. Ler simply means "the sea(s)", and the mythological character of Lír is best known as father of "The Children of Lír" and of the sea-loving trickster, Manannán Mac Lír, for whom the Isle of Man is named. Sinann, then, is grand-daughter of The Sea.
For more on Manannán, see "Rowing Around Immráma 12 – In Search of Manannán."

*Lodán* is a curious name. The -án suffix is common as a diminutive (i.e. x-án = "little x"), and is therefore a "fond" form in many personal names e.g. Cíarán = "(my) little Dark One". However, the word lodán, while straight-forwardly meaning a "little pool", tends to mean a puddle of filth, even excrement!

Taking lod as the root of the name makes more sense. It is quite clear that this means a pool, a lake or a boundary of water. Interestingly, more than one source glosses lod as the place where sruth-aili emerges. On the surface, it seems prosaic to say that a stream (sruth) comes from a pool. But there is an advanced poetic metre called sruth di aill, "stream from a cliff / stone", so the association of lod as a source-pool or well-spring of sruth di aill may well have been intended by the poets. As such, Lodán becomes a fitting father for Sinann.
There is also a place known as Lodan which Hogan (in his "Onomasticon Goedelicum") places between the River Mulkern and Loch Gur in Co. Limerick – close to, if not encompassed by, the area of Luachair [see below].

<u>Luchair:</u> Literally "reeds" or "rushes", a common element in placenames. However, Lu[a]chair on its own refers to an area of North Kerry and Cork around Sliabh Luachra. Another meaning for lúachair is "brilliance" or "brightness", and poetic descriptions of the region of Lúachair often include words invoking fire, light, shining etc.

*Segas*: This is most often cited as the name of an otherworld well that gives rise to the great rivers of Ireland. Vendryes translates it as "forest", but another word, séganda, means "magnificent" or "skillful", and ségannacht means "outstanding skill" or "ingenuity". Other possible roots are seg, which refers to "milk" (see Sinann below), and sed / seg, meaning "vigour" or "strength". It may also relate to saigid, "to seek, strive or aim for", or ségda, meaning "lucky, happy, fortuitous". Then there's écosc, an "appearance or characteristic" by which someone or something is known for what they are.

However, given the context and characteristics of the well called Segas, [Isolde] would like to suggest its origin as éces; "scholar, poet, knowledgeable one". This word seems to share a root with do-éici, and so might mean "a seer" in origin, just as file originally means "seer". That poets and scholars could "see" relates to the immas forosna, which was a way of foretelling events through improvised poetry. It is this immas that Sinann sought from Segas, and this cluster of meanings seems most satisfying. You can read more about Imbás and Poets here: https://storyarchaeology.com/2012/06/21/imbas/ .

*Sinann:* In keeping with many names that use the –ann ending, I would propose an earlier form of *Siniu (sinann would then be genetive: sruth Sinann = "stream of Siniu"). There is the word sín, meaning a "storm" or "bad weather", used in the last line of the second poem. This may be the Sin of the story with Muirchertach, whereby he comments on the bad weather, and she replies, "why do you speak my name?".
However, for our river, I feel the most likely root is with sine, "a teat or breast". Indeed, there are stories with female characters (notably, a lake monster in Loch Rudraige!) called Sinech. The name could derive from sen, "old", but this does not seem to fit our story, which continually describes her as young.

*Sín Morainn:* This is usually translated "Morainn's Collar". There is a tradition that this was a chain or yoke that he wore, but it could also be an epistle used like a talisman. An entry in Cóir Anmann, The Fitness of Names, mentions it:

> *Feradhach Fechtnach .i. ar feachtnaighi a flaithiusa for Eirinn, nam feachtnach .i. fírén .i. ar fírinne a flaitha adubhradh Feradhach Fechtnach fris, ar is a n-aimsir Fearadhaigh robói an Idh Mhorainn & Morann feissin. Issí an Idh sin Morainn nóghléedh fírinne do chách. Iss aire sin adubradh ind aghnomen fris*
>
> *Feradach Fechtnach 'the Righteous', because of the fechtnaige 'righteousness' of his reign over Erin. For fechtnach means righteous: that is, for the truth of his reign he was called Feradach Fechtnach. For in his time was the Collar of Morann and Morann himself. It is that Collar of Morann that used to declare truth to every one. Therefore the epithet (Fechtnach) was given to him (Feradach).*

Morann's Three Collars (or three stories of the same collar!) feature in Cormac's Cup and The Twelve Ordeals – W Stokes, the text we examined when discussing The Shocking Revelations Concerning King Cormac Mac Airt in Rowing Around Immráma episode 8 [of *Story Archaeology* podcasts].

**sí in moirenn:** The principle definition of moirenn [muirenn] given in eDIL is that of "spear". However, there seems to be another meaning of "sea" or "waves", perhaps from imagining the waves of the sea as spears hurled from the depths. This seems a more likely translation in the context of Sinann, although the original audience of the poem may have understood both meanings as present.

Appendix C2: O'Curry's telling of Sinann's story and Ó Brolcháin Carmody's critical notes

Here we re-publish Eugene O'Curry's telling of the story of Sinann [O'Curry,1873]. This is followed by accompanying notes by Isolde Ó Brolcháin Carmody. We label these notes [IOB] to identify the part of O'Curry's text to which they pertain.

Eugene O'Curry's telling:

> The third piece of O'Lothchain's composition is a poem of sixteen stanzas, or fifty-six lines, on the origin of the name of the Sinann, now the River Shannon. This poem begins: "The noble name of Sinainn seek ye from me; Its bare recital would not be pleasant, Not alike now are its action and noise As when Sinann herself was free and alive"…[OCB1] It is shortly as follows:

Sinann was the daughter of the learned Lodan, who was the son of Lear, the great sea-king of the Tuatha Dé Danann colony of Erinn, from whose son and successor Manannan the Isle of Man derives its name and ancient celebrity. In those very early times there was a certain mystical fountain which was called Connlas Well, (situated, so far as we can gather, in Lower Ormond). [OCB2] As to who this Connla was, from whom the well had its name, we are not told; but the well itself appears to have been regarded as another Helicon by the ancient Irish poets. Over this well there grew, according to the legend, nine beautiful mystical hazel-trees, which annually sent forth their blossoms and fruits simultaneously. The nuts were of the richest crimson colour, and teemed with the knowledge of all that was refined in literature, poetry, and art. No sooner, however, were the beautiful nuts produced on the trees, than they always dropped into the well, raising by their fall a succession of shining red bubbles. Now during this time the water was always full of salmon; and no sooner did the bubbles appear than these salmon darted to the surface and eat the nuts, after which they made their way to the river. [OCB3] The eating of the nuts produced brilliant crimson spots on the bellies of these salmon; and to catch and eat these salmon became an object of more than mere gastronomic interest among those who were anxious to become distinguished in the arts and in literature without being at the pains and delay of long study; for the fish was supposed to have become filled with the knowledge which was contained in the nuts, which, it was believed, would be transferred in full to those who had the good fortune to catch and eat them. Such a salmon was, on that account, called the Eo Feasa, or "Salmon of Knowledge"; and it is to such a salmon that we sometimes meet a reference among our old poets, where, when speaking of objects which they pretend to be above description, they say, "unless they had eaten of the salmon of knowledge, they could not do it justice"….

To proceed, however, with the legend of the Shannon: It was forbidden to women to come within the precincts of Connla's wonderful well; [OCB4] but the beautiful lady Sinann, who possessed above every maiden of her time all the accomplishments of her sex, longed to have also those more solid and masculine acquirements which were accessible at Connla's well to the other sex only. [OCB5] To possess herself of these she went secretly [OCB6] to the mystical fountain; but as soon as she approached its brink, the waters rose up violently, burst forth over its banks, and rushed towards the great river now called the Shannon, overwhelming the lady Sinann in their course, whose dead body was carried down by the torrent, and at last cast up on the land at the confluence of the two streams. After this the well became dry for ever; [OCB7] and the stream which issued from it was that originally known by the name of the lady Sinann or Shannon; but having fallen into that great succession of lakes which runs nearly through the centre of Ireland, the course of lakes subsequently appropriated the name to itself, which it still retains, whilst the original stream is now unknown. [OCB8] The original Sinann is, however, believed to have fallen into the present Shannon, near the head of Loch Dearg, not far from Portumna.

Critical notes by Isolde Ó Brolcháin Carmody:

[OBC1]: This is the poem published as "Sinann I" in Gwynn's Metrical Dindshenchas, Vol 3, poem 53, p. 286 ff [Gwynn, 1906].

[OBC2]: Lower Ormond is South-East Munster, Waterford or Wexford. I don't know O'Curry's basis for this. The poem clearly places the well of Condla under the sea.

[OBC3]: The sequence of nuts-bubbles-salmon is a bit confused: in the poem, the bubbles are formed by the juice squeezed from the nuts as the salmon eat them.

[OBC4]: It does not say in either Dindshenchas poem that it was forbidden for women to approach the well.

[OBC5]: I cannot find such gender distinctions in either poem. Neither were the learned professions "forbidden" to women in ancient Ireland, although professional women usually had lower status than their male counterparts.

[OBC6]: The poem says that she had the idea to visit the well one night – it says nothing of it being a secret, and in fact states that she brought the finest of her household with her on this supposedly "secret" journey.

[OBC7]: I don't know from where O'Curry gets this information of the well drying up after Sinann's visit.

[OBC8]: Again, I think there is some confusion about the "streams" mentioned in the poem, of which one is the Shannon. Perhaps O'Curry, believing the well to be somewhere in Lower Ormond (South-East Munster) imagines another river reaching from there to the "current" Shannon at Lough Derg, but that this stream is somehow now lost. If he was looking for a river or stream in that region as a candidate, he might have settled on the river Nenagh, which is locally known as "Sinann". This is in North Tipperary, not too far from Lough Derg, where O'Curry says the body of Sinann was washed up near Portumna.